\documentclass[%
 preprint, 
superscriptaddress,
 amsmath,amssymb,
 aps, physrev,
prb,
]{revtex4-2}
\usepackage{xcolor}
\usepackage{graphicx}
\usepackage{dcolumn}
\usepackage{bm}
\usepackage{multirow}

\begin{document}

\preprint{APS/123-QED}

\title{\textbf{Orbital-resolved superexchange and topological magnon bands in MXene Fe$_2$C} 
}%

\author{Qiang Gao}
\email{waveflying@163.com} 
\affiliation{School of Science, Shenyang University of Technology, No. 111 Shenliao West
Road, Shenyang 110870,  China}
\affiliation{Shenyang National Laboratory for Materials Science, Institute of Metal Research, Chinese Academy of Sciences, Shenyang 110016,  China} 

\author{Wei Wang}
\affiliation{School of Science, Shenyang University of Technology, No. 111 Shenliao West
Road, Shenyang 110870,   China} 

\author{Shaojin Qi}
\affiliation{School of Science, Shenyang University of Technology, No. 111 Shenliao West
Road, Shenyang 110870,   China}

\author{Guimei Shi}
\affiliation{School of Science, Shenyang University of Technology, No. 111 Shenliao West
Road, Shenyang 110870,  China}

\author{Anbang Guo}
\affiliation{School of Science, Shenyang University of Technology, No. 111 Shenliao West
Road, Shenyang 110870,  China}

\author{Xin Jin}
\thanks{Corresponding author. Email:  jinxin931215@163.com}
\affiliation{College of Physics and Electronic Engineering, Chongqing Normal University,
No. 55 Daxue City South Road, Chongqing 401331, China}

\author{Chen Shen}
\affiliation{Suzhou Laboratory, Suzhou, China}

\author{Xing-Qiu Chen}
\thanks{Corresponding author. Email:  xingqiu.chen@imr.ac.cn} 
\affiliation{Shenyang National Laboratory for Materials Science, Institute of Metal Research, Chinese Academy of Sciences, Shenyang 110016,  China}



\date{\today}

\begin{abstract}
Magnetic MXenes are promising candidates for future spintronics due to their intriguing properties. Nevertheless,   the   topological properties of magnon bands   and superexchange mechanism  remain relatively underexplored.    In this work, we have calculated the spin wave dispersions for MXene Fe$_2$C  using the linear spin wave method, with the exchange coupling parameters  obtained from first-principles calculations. Owing to the staggered stacking of two Fe sub-lattices, 
the interlayer exchange coupling lifts the degeneracy   of the ferromagnetic magnon modes.   A Dirac point is identified in the magnon bands at the $K$ point. 
The  topological properties of magnon bands, including  Berry curvature,   valley Chern number and  edge states,  are further computed by means of a two-band model.  
We also derive an effective Hamiltonian to explain the magnonic topology, which is protected by the  $C_{3v}$ rotational symmetry.
To understand the microscopic origin of the exchange couplings that govern the magnon bands, we then construct a refined    model   integrating perturbation theory and the tight-binding   approach.
The ferromagnetic superexchange coupling  is mediated by the virtual hopping between the d orbitals of two distinct Fe sites  via a   pair of orthogonal p orbitals at the bridging C site.
In detail, the intralayer exchange coupling is   governed  by the in-plane $p_x$ and $p_y$ orbitals, while the out-of-plane $p_z$ orbital is responsible for the interlayer exchange coupling. 
Our results establish a direct link between the orbital-resolved anisotropic  superexchange and topological properties of the magnon bands in  magnetic MXene Fe$_2$C, providing  theoretical guidance for future  spintronics  applications.

\end{abstract}

\maketitle


\section{Introduction}
Since the landmark discovery of Graphene, two-dimensional (2D) materials have been  attracting substantial attention in the fields of quantum,  green energy, and advanced information technologies, because of their fascinating properties and significant application potential~\cite{novoselov2004electric,geim2007rise}.  
With decades of extensive research efforts, a vast library of  2D materials has been theoretically predicted and experimentally fabricated~\cite{Pei2026,ying2025facing}. 
In particular, 2D magnetic materials are regarded as promising candidates for the next-generation spintronics, owing to their capability for multi-field tunable electronic design~\cite{gong2019two,jia2025spintronic}, {\it i.e.}  van der Waals heterostructures, magneto-piezoelectric coupling, magneto-electric coupling, and strain-mediated magnetic responses.
However, despite intensive research efforts, 2D magnetic materials remain scarce in comparison to the vastly broader class of nonmagnetic 2D systems.  Representative examples include   CrI$_3$~\cite{jiang2018controlling}, Cr$_2$Ge$_2$Te$_6$~\cite{gong2017discovery}, MnBi$_2$Te$_4$~\cite{zhang2019topological},  TMPS$_3$ (TM=Fe and 
Mn)~\cite{li2014half,li2013coupling}, CrTe$_x$ systems~\cite{zhang2025advances}, and so on.  
Two fundamental challenges hinder the  exploration of 2D magnets. One typical problem is that the critical temperatures of most experimentally realized cases are  far  below room temperature, because the  dimensional crossover will  reduce  the coordination number of exchange coupling  and  weaken wavefunction overlapping between magnetic orbitals~\cite{gong2019two,jiang-2021-study,Igor-2019-micorscopic}. 
Furthermore,  according to Mermin-Wagner theorem~\cite{mermin1966absence},
the long-range magnetic order in low-dimensional systems can only be stabilized against thermal fluctuations at finite temperature  by opening a gap in the spin-wave spectrum via magneto-crystalline anisotropy energy (MAE), which  originates primarily from  spin-orbit coupling (SOC). 
To this end, for the practical applications, it is crucial to design 2D magnets that simultaneously combine a relatively high Curie temperature and  a sizable spin-wave gap.

In the field of 2D materials,  the MXene compounds have formed a large family due to  the diverse   chemical composition~\cite{naguib2011two,naguib2023two,ZHOU20242776}.
In particular,  the MXene nano-sheet can be readily fabricated  by   etching  the corresponding bulk  MAX-phase compound.
At the same time,  some  MXenes  have also   been  predicted to realize 2D magnetism with multifunctional applications. For instance, theoretical calculations indicate that the MXene  monolayer V$_2$N  can become  a spin-gapless semiconductor (SGS) under a biaxial strain~\cite{gao2016monolayer}. Similarly, the Janus MXene TaFeC is predicted to be a new type of ferromagnetic (FM) SGS with a MAE as large as  0.589 meV/f.u.~\cite{gao2025first}  
Density functional theory (DFT) calculations find that
electric field will induce Zeeman-like splitting into band structures and the magneto-optical Kerr effect in the functionalized anti-ferromagnetic (AFM) MXene Cr$_2$CCl$_2$.
Based on high-throughput screening,  Gao et al.~\cite{gao2020magnetic}  predict that i-MXene compounds (X$_{2/3}$Fe$_{1/3}$)$_2$C (X=Ta, Zr, and Hf)  have MAE values as high as 0.86 meV/f.u.,  0.74 meV/f.u., and 1.39 meV/f.u., respectively. 
The subsequent studies have also revealed that   oxygen functional group modified MXene (Ta$_{2/3}$Fe$_{1/3}$)$_2$C  is predicted to be a type-I multiferroic material with a moderate polarization up to  about 12.33  $\mu$C/cm$^2$ along $a$-axis~\cite{zhao2021multiferroic}. 
Theoretical calculations find that the chemical group modified Janus MXene VYNF$_2$ displays AFM order
with flat robust bands, valley splitting, and
an out-of-plane piezoelectric response of 0.14 pm $\cdot$V$^{-1}$~\cite{gao2025first-VYNF2}.
Especially,  the pristine MXenes tend to have remarkably high magnetic phase transition temperatures. For instance,  DFT calculations suggest that  the Curie temperatures of Cr$_2$C~\cite{si2015half,sun2020tunable}  and    Fe$_2$C~\cite{YUE2017164,agapov2024mxene,LOU2022169959}   are  respectively as high as 672 K and  510  K. However, the MAE values for such  two monolayers are just  0.169 meV with an out-of-plane magnetization~\cite{akgencc2021tuning} and 0.228 meV with an in-plane magnetization~\cite{YUE2017164}.  
In particular,  the lanthanide based MXenes have been recently fabricated  by intercalating layered halides as van der Waals building blocks~\cite{fang2026semiconducting}, providing new opportunities for magnetic MXenes.
Apparently,  magnetic MXenes are promising candidates for future spintronic applications.

Recent studies have indicated that the magnon bands  can also  host  topological non-trivial phase~\cite{onose2010observation,katsura2010theory,owerre2016first}, in analogy to topology in electronic band structures. 
For instance, a toy model study~\cite{ortmanns2021magnon} finds that the magnon band topology emerges in AB-stacking bilayers of the honeycomb sub-lattice, 
where the interlayer exchange coupling hybridizes the magnon excitations of the two magnetic layers hence induces Dirac‑point degeneracy at  the $K$ point. 
For CrI$_3$, the exchange coupling and consequently the magnon band topology can be manipulated by the external stimuli (strain or magnetic field), where the topological non-trivial phase can be switched into a topological trivial one by strain~\cite{prm-2023-soenen}. By forming a vdW heterostructure with different
(nonmagnetic) hexagonal 2D materials or substrates on either side, the off-diagonal exchange coupling can be induced into CrI$_3$, leading to magnon valley  Hall effect~\cite{sacoto-2020-prb}. It is reported that magnon valley  Hall effect can also be induced by surface acoustic waves driven by the spatially modulated exchange coupling~\cite{faizee2026magnon}.  
Clearly, the exchange coupling  plays an essential role in the topology of magnon bands. 
Since the topology is determined by the magnetic interactions, it is of importance to understand the microscopic origin of the exchange couplings.
In 2D magnets, the  underlying mechanism is superexchange mediated by ligand bridging, where the superexchange coupling can usually be explained by the  Goodenough-Kanamori-Anderson (GKA) rules~\cite{Goodenough-1955,KANAMORI195987,Anderson-1950}.
For instance,   the superexchange path Cr-Bi-Cr is responsible for the enhanced Curie temperature (340 K) in the heterostructure CrSBr/Bi,  compared with 179 K Curie temperature in the pristine monolayer CrSBr~\cite{zhou-2205}. 
Both lattice size and ligand electronegativity can modulate the superexchange strength as shown in the monolayers 2H-FeXY (X,Y= Cl, Br, I)~\cite{qvwf-vhc6}. 
In this context, the superexchange mechanism provides  crucial insight into  the topology of magnon bands in 2D magnets, since  superexchange is always involved in the magnetic interactions in such materials.

In this work, we have studied the topological properties of the magnon band and the superexchange mechanism for the magnetic MXene Fe$_2$C.
The magnetic interactions for Fe$_2$C are relatively simple, consisting solely of  intralayer and interlayer exchange couplings~\cite{YUE2017164,agapov2024mxene,LOU2022169959}.
In particular, the presence of the interlayer exchange coupling can induce off-diagonal  terms into the Hamiltonian, lifting  degeneracy of ferromagnetic spin‑wave modes.  
Clearly, the MXene Fe$_2$C is  an ideal and minimal platform to study topology of magnon bands within two-band model.
Moreover, the superexchange coupling of Fe$_2$C  can be simplified and well interpreted  based on perturbation theory, 
without the confounding effects of the complicated magnetic interactions.
To be specific, we have calculated the spin wave dispersion for MXene Fe$_2$C with a special focus on the topological properties using a two-band model, and provided a detailed analysis of the underlying super‑exchange coupling mechanism based on perturbation theory.

\section{Computational details and model}
\subsection{First-principles calculations}
The density functional theory (DFT) calculations are carried out by  projector augmented wave method in the Vienna
{\it ab initio} simulation package (VASP)~\cite{kresse_efficient_1996,kresse_ultrasoft_1999}   with the assistant of   VASPKIT code~\cite{wang2021vaspkit}.
The calculations are based on the exchange-correlation functional of the generalized gradient
approximation (GGA) with the Perdew-Burke-Ernzerhof (PBE)~\cite{perdew1996generalized}, where the 
valence electron configurations of Fe and C are treated as 3d$^7$4s$^1$ and 2s$^2$2p$^2$. 
To avoid spurious interactions between the periodically replicated monolayer surfaces, a 20 \AA \space thick vacuum layer is introduced along the $z$-axis to isolate the 2D structure.
The plane wave energy cutoff is set as 500 eV.
The  criteria  of force and total energy are respectively set as 0.001 eV/\AA \space  and 10$^{-6}$ eV  in all calculations.
 The whole Brillouin zone (BZ) is sampled with a $k$-mesh of 21$\times$21$\times$1.  
To derive the mechanism behind the superexchange coupling, symmetry-adapted  Wannier functions are constructed based on the  DFT calculations of full-potential local-orbital minimum-basis code (FPLO)~\cite{koepernik1999full,opahle1999full} in an automated method~\cite{zhang2018high}.
A k$-$mesh of 21$\times$21$\times$1 is set for the self-consistent calculations in FPLO code.
The local density of states is obtained by the obtained Wannier functions.

\subsection{Linear Spin-Wave Theory}

\begin{figure*}
    \centering
    \includegraphics[width=0.88\linewidth]{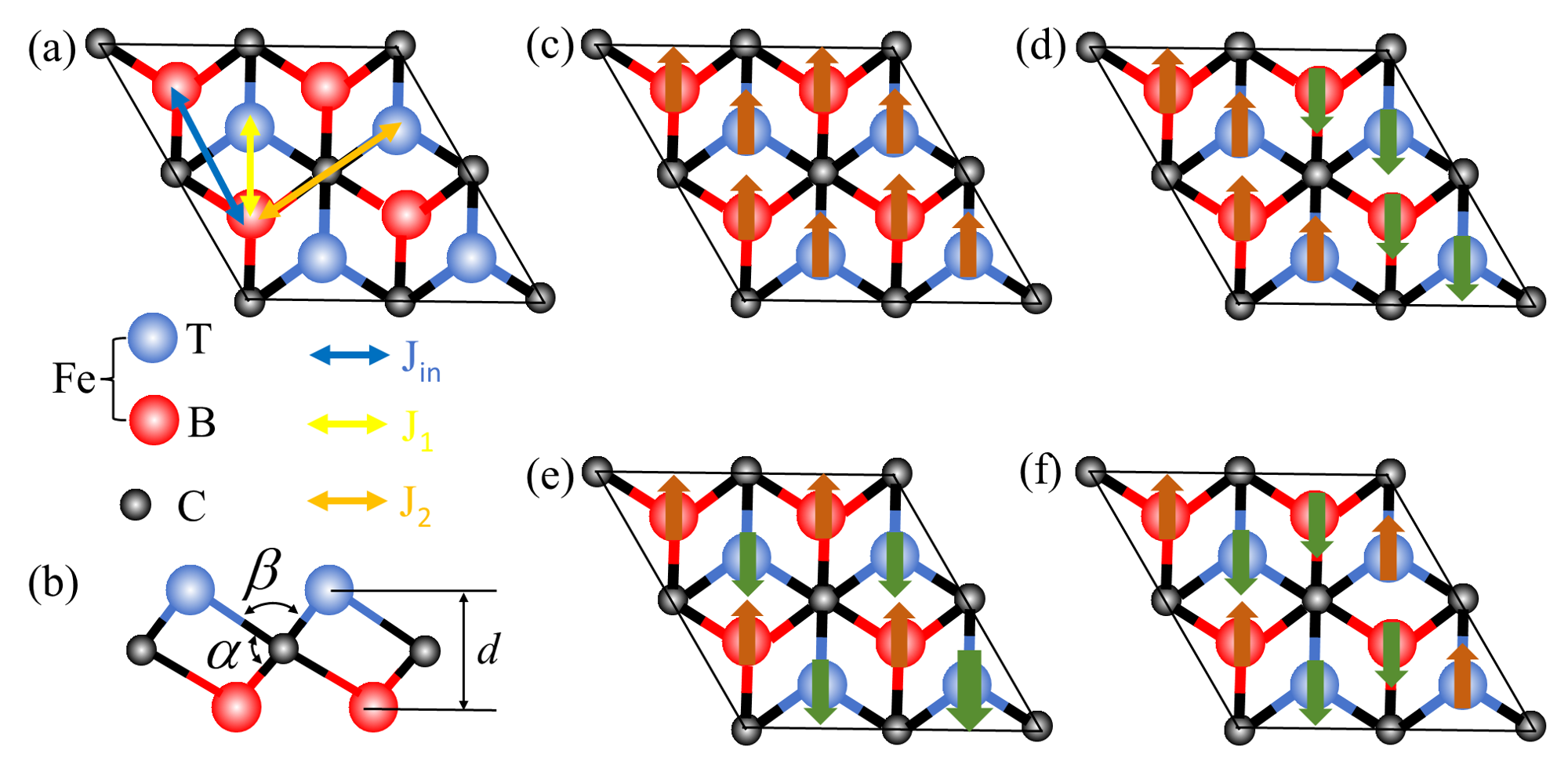}
    \caption{The top (a) and side (b) views for the structure of MXene Fe$_2$C. (c), (d), (e), and (f) represents FM, AFM-A, AFM-C, and AFM-G spin configuration, respectively.  The bottom (B) and top (T) layers of Fe are distinguished by the red  and blue atoms.
    In (a), the blue, yellow and orange double-headed arrows denote the interactions of the intralayer exchange coupling, nearest neighbour interlayer exchange coupling and next nearest neighbour interlayer exchange coupling.
   In (c)--(f), the  brown and green arrows denote the spin parallel to up and down directions.}
    \label{fig-1}
\end{figure*}

To derive the spin wave, we begin with the structure for MXene Fe$_2$C.
As shown in Fig.~\ref{fig-1} (a), the Fe atoms are  vertically staggered on the top and bottom layers of Carbon, leading two  interpenetrating triangular sublattices. In the present study, we have considered the nearest neighbouring (NN)  intralayer exchange coupling for the bottom and top layers of Fe together with the NN and next nearest neighbouring (NNN) interlayer exchange coupling between the top and bottom layers of Fe. The coordination number of the NN intralayer exchange coupling is $Z_{in}=$6, while that of the NN and NNN interlayer exchange coupling are both $Z_{ou}=$3.  Accordingly, the structure factors for such three types of exchange coupling are respectively expressed by 
\begin{equation}\label{structure-factor}
\begin{cases}
\gamma_{in} (\boldsymbol{k}) = \frac{2}{Z_{in}} \left[ \cos k_xa+2\cos\frac{k_xa}{2}\cos\frac{\sqrt{3}k_ya}{2} \right]  \\
\gamma_{1} (\boldsymbol{k})   = \frac{e^{i k_z d}}{Z_{ou}} \left[ 2 \cos\left( \frac{k_x a}{2} \right) e^{i \frac{\sqrt{3} k_y a}{6}} + e^{-i \frac{\sqrt{3} k_y a}{3}} \right]   \\
\gamma_{2} (\boldsymbol{k})   = \frac{e^{i k_z d}}{Z_{ou}} \left[ 2 \cos\left( k_x a \right) e^{-i \frac{\sqrt{3} k_y a}{3}} + e^{i \frac{2\sqrt{3} k_y a}{3}} \right]
\end{cases}.
\end{equation}
Here, $a$ and $d$ denote the in-plane lattice constant and the distance between the top and bottom layers of Fe. It is noticed that $e^{i k_z d}=1$ since the out-of-plane ($z$) component of the wave vector vanishes  in  2D materials.
Notably, the structure factor $\gamma_{in}$ of the NN intralayer exchange coupling is a real  function in the whole Brillouin zone, which is the reflection of the six-fold rotational symmetry of intralayer spin waves. On the other hand, the structure factors $\gamma_{1} (\boldsymbol{k})$ and 
 $\gamma_{2} (\boldsymbol{k})$ of the NN and NNN interlayer exchange coupling are  complex numbers  except for the high-symmetry points (K, $\Gamma$ and M), leading to  the hybridized mode induced by interlayer exchange coupling.

Considering the in-plane magnetic anisotropy, the  Heisenberg Hamiltonian for the present  system is described by
\begin{equation}\label{ha1}
\small
\begin{split}
H=&- J_{in}\sum_{\lambda\in\left\{T,B\right\},\mathbf{r}_{i}, \mathbf{\rho}_{in}}{\mathbf{S}_{\mathbf{r}_{i},\lambda} \cdot \mathbf{S}_{\mathbf{r}_{i}+\mathbf{\rho}_{in}, \lambda}}-\sum_{l\in\{1,2\},\mathbf{r}_{i},\rho_{ou}} J_{l}\mathbf{S}_{\mathbf{r}_{i},T} \cdot \mathbf{S}_{\mathbf{r}_{i}+\mathbf{\rho}_{ou},B}\\
&-K\sum_{\mathbf{r}_{i},\lambda\in\left\{T,B\right\}}(S^z_{\mathbf{r}_{i},\lambda})^2   
\end{split}.
\normalsize
\end{equation}
where $\mathbf{\rho}$, $S$  and $K$ denote  the lattice displacement, quantum spin number and  magnetic anisotropy energy for each Fe, respectively. In the whole manuscript, the subscript  ``T'', ``B'', ``in'' and ``ou'' mark the  top layer of Fe,   bottom layer of Fe,  intralayer related term, and interlayer related term, respectively.
J$_{in}$ is the NN intralayer exchange coupling parameter, while J$_1$ and J$_2$ denote the NN and NNN interlayer exchange coupling parameters between the top and bottom layers of Fe, respectively.  Notably, in the intralayer summation, the factor   $\frac{1}{2}$ has been absorbed into the definition of $J_{in}$ to avoid double counting.
In Eq.~\eqref{ha1}, the first term   is the contribution from the intralayer exchange coupling.
The second term describes the contribution from the NN and NNN neighbouring interlayer exchange coupling between the top and bottom layers of Fe. The third term is the magnetic anisotropy energy. 
Notably,  positive and negative signs for exchange coupling parameters correspond to the ferromagnetic (FM) and antiferromagnetic (AFM) types, respectively. 
As to K, a positive (negative) sign suggests the magnetization prefers an out-of-plane (in-plane) direction. 
The linear spin wave theory~\cite{Toth_2015} is adopted to   derive the magnon spectra under the assumption of weak excitation at low temperature.   For Fe$_2$C,  the considered exchange couplings are all ferromagnetic, including the intralayer exchange couplings of the bottom and the top  layers of Fe as well as the interlayer exchange coupling.

Firstly, we will focus on the intralayer exchange coupling term  in Eq.~\eqref{ha1}. 
Using Holstein-Primakoff (HP) transformation~\cite{holstein1940field},  the spin operators of bottom  and the top sublattices are respectively written by
\begin{equation}\label{hpa1}
S^+_{B,i}=(\sqrt{2S-a_i^\dagger a_i})a_i,\space S^-_{B,i}=a^\dagger_i(\sqrt{2S-a_i^\dagger a_i}), \space S_{B,i}^z=S-a_i^\dagger a_i
\end{equation}
\begin{equation}\label{hpa2}
S^+_{T,j}=(\sqrt{2S-b_j^\dagger b_j})b_j,\space S^-_{T,j}=b^\dagger_j(\sqrt{2S-b_j^\dagger b_j}), \space S_{T,j}^z=S-b_j^\dagger b_j.
\end{equation}
Here, $S^\pm_{B/T,i}$ and $S^z_{B/T,j}$ denote the spin ladder operators and projected spin operator in $z-$axis for the  $i$-th site in the bottom of Fe and $j-$th site in the top layer of Fe, respectively.   $a_i^\dagger$ and $a_i$  ($b_j^\dagger$ and $b_j$) are the Boson spin creation and annihilation operators. 
Hereafter, the intralayer exchange coupling term in Eq.~\eqref{ha1} for the bottom and top layers of Fe are rewritten by 
\begin{equation}\label{hpa3}
\begin{split}
H_B &= -J_{in} \sum\limits_{i,\rho} \Bigg\{ \left(S - a_i^\dagger a_i\right)\left(S - a_{i+\rho}^\dagger a_{i+\rho}\right) \\
  &\quad + \frac{1}{2} \sqrt{2S - a_i^\dagger a_i} \, a_i a_{i+\rho}^\dagger \sqrt{2S - a_{i+\rho}^\dagger a_{i+\rho}} \\
  &\quad + \frac{1}{2} a_i^\dagger \sqrt{2S - a_i^\dagger a_i} \sqrt{2S - a_{i+\rho}^\dagger a_{i+\rho}} \, a_{i+\rho}
\Bigg\}
\end{split}
\end{equation}
and
\begin{equation}\label{hpa4-kws}
\begin{split}
H_T &= -J_{in} \sum\limits_{j,\rho} \{ \left(S - b_j^\dagger b_j\right)\left(S - b_{j+\rho}^\dagger b_{j+\rho}\right) \\
  &\quad + \frac{1}{2} \sqrt{2S - b_j^\dagger b_j} \, b_j b_{j+\rho}^\dagger \sqrt{2S - b_{j+\rho}^\dagger b_{j+\rho}} \\
  &\quad + \frac{1}{2} b_j^\dagger \sqrt{2S - b_j^\dagger b_j} \sqrt{2S - b_{j+\rho}^\dagger b_{j+\rho}} \, b_{j+\rho}
\}.
\end{split}
\end{equation}

Given that the number of the excited magnon is  sufficiently low, the magnon-magnon interaction
can be neglected~\cite{halilov1997magnon}. So, we can get the approximations of
$\sqrt{2S - a_i^\dagger a_i}\approx \sqrt{2S - b_i^\dagger b_i}\approx \sqrt{2S}$.
The ladder operators in Eqs.~\eqref{hpa1} and~\eqref{hpa2} are approximated as
\begin{equation}\label{hpa5}
S^+_{B,i}\approx \sqrt{2S}\,a_i,\quad S^-_{B,i} \approx\sqrt{2S}\, a^\dagger_i,
\end{equation}
\begin{equation}\label{hpa6}
S^+_{T,i} \approx \sqrt{2S}b_i,\quad S^-_{T,i} \approx \sqrt{2S} b^\dagger_i.
\end{equation}
Applying Fourier transformation, the Boson operators can be rewritten in the collective coordinates in momentum space as
\begin{equation}\label{reciprocal-cange1}
    a_{i} = \frac{1}{\sqrt{N}} \sum_{\boldsymbol{k}} e^{i\boldsymbol{k}\cdot\boldsymbol{r}_i} a_k, \quad a_{i}^\dagger = \frac{1}{\sqrt{N}} \sum_{\boldsymbol{k}} e^{-i\boldsymbol{k}\cdot\boldsymbol{r}_i} a_k^\dagger ,
\end{equation}
\begin{equation}\label{reciprocal-cange2}
    b_{j} = \frac{1}{\sqrt{N}} \sum_{\boldsymbol{k}} e^{i\boldsymbol{k}\cdot\boldsymbol{r}_j} b_k, \quad b_{j}^\dagger = \frac{1}{\sqrt{N}} \sum_{\boldsymbol{k}} e^{-i\boldsymbol{k}\cdot\boldsymbol{r}_j} b_k^\dagger .
\end{equation}
Therefore, the intralayer exchange coupling term (the first term in Eq.~\eqref{ha1}) for  the Heisenberg Hamiltonian can be reformulated by
\begin{equation}\label{hpa4}
\begin{split}
  H_{in} &=-2NZ_{in}J_{in}S^2+ 2Z_{in}J_{in}S\left[1-\gamma_{in} (\boldsymbol{k}) \right]     \sum_k \left(   a^\dagger_k a_k+     b^\dagger_k b_k\right)\\
    &=-2NZ_{in}J_{in}S^2+ \sum_k  \left[ \hbar \omega_B (\boldsymbol{k})a^\dagger_k a_k +   \hbar \omega_T (\boldsymbol{k})b^\dagger_k b_k     \right] 
\end{split},
\end{equation}
where  the $\hbar \omega_B (\boldsymbol{k})= \hbar \omega_T (\boldsymbol{k})=2Z_{in}J_{in}S\left[1-\gamma_{in} (\boldsymbol{k}) \right] $, representing the degenerated spin wave of the bottom and top layers of Fe.
Here the constant term ($-2NZ_{in}J_{in}S^2$) will be summarized into the energy of the ground state in the final Hamiltonian.

Secondly, the interlayer exchange term in Eq.~\eqref{ha1} is explicated also by using the linear spin wave theory. 
In terms of the Boson spin operators (Eqs.~\eqref{hpa1} and~\eqref{hpa2}), the interlayer term in Heisenberg Hamiltonian is expressed by 
\begin{equation}\label{interlayer-ham}
\small
\begin{split}
H_{ou}=&-J_1\sum\limits_{i,\rho^{(1)}} 
\left\{ S^z_{B,i}S^z_{T,i+\rho^{(1)}}+\frac{1}{2} \left[  S^+_{B,i} S^-_{T,i+\rho^{(1)}} +  S^-_{B,i} S^+_{T,i+\rho^{(1)}}  \right]\right\} \\
&-J_2\sum\limits_{i,\rho^{(2)}} 
\left\{ S^z_{B,i}S^z_{T,i+\rho^{(2)}}+\frac{1}{2} \left[  S^+_{B,i} S^-_{T,i+\rho^{(2)}} +  S^-_{B,i} S^+_{T,i+\rho^{(2)}}  \right]\right\} 
\end{split},
\normalsize
\end{equation}
where  $\rho^{(1)}$ and $\rho^{(2)}$ are respectively the displacements of the NN and NNN interlayer exchange coupling pairs between the bottom and top layers of Fe.
As detailed  above, we have already obtained the approximated expressions of  the ladder operators in Eqs.~\eqref{hpa5} and~\eqref{hpa6} for the bottom and top layers of Fe, by using linear spin wave theory.
So, the interlayer  Heisenberg Hamiltonian in Eq.~\eqref{interlayer-ham} can be reformulated by  
\begin{equation}\label{interlayer-lswt}
\small
\begin{split}
H_{ou}=&-NZ_{ou}(J_1+J_2)S^2 \\
&+J_1S\sum\limits_{i,\rho^{(1)}}\left\{ a^\dagger_ia_i+b^\dagger_{i+\rho^{(1)}}b_{i+\rho^{(1)}} -\left[a^\dagger_ib_{i+\rho^{(1)}}+ a_ib^\dagger_{i+\rho^{(1)}} \right]  \right\}\\
&+J_2S\sum\limits_{i,\rho^{(2)}}\left\{ a^\dagger_ia_i+b^\dagger_{i+\rho^{(2)}}b_{i+\rho^{(2)}} -\left[a^\dagger_ib_{i+\rho^{(2)}}+ a_ib^\dagger_{i+\rho^{(2)}} \right]  \right\}
\end{split}.
\normalsize
\end{equation}
In the above, we have neglected the fourth order term of $a^\dagger_ia_ib^\dagger_{i+\rho^{(1/2)}}b_{i+\rho^{(1/2)}}$, since the magnon-magnon interaction is very weak at low temperature.
Substituting  Eqs.~\eqref{reciprocal-cange1} and~\eqref{reciprocal-cange2}, the above Hamiltonian is rewritten in terms of the  Boson operators in the reciprocal space by

\begin{equation}\label{interlayer-ha-recp}
\begin{split}
H_{ou}=&-NZ_{ou}\left(J_1+J_2\right)S^2 +Z_{ou}\left(J_1+J_2\right) S \sum_{\boldsymbol{k}}\left(a_{\boldsymbol{k}}^\dagger a_{\boldsymbol{k}}+b_{\boldsymbol{k}}^\dagger b_{\boldsymbol{k}}\right) \\
&-Z_{ou}S\sum_{\boldsymbol{k}}\{\left[J_1\gamma_1(\boldsymbol{k})+J_2\gamma_2(\boldsymbol{k})\right] a_{\boldsymbol{k}}^\dagger b_{\boldsymbol{k}} + h.c. \}
\end{split}.
\end{equation}

Lastly, we will treat the magnetic anisotropy term. 
Based on HP transformation, the magnetic anisotropy term can be expressed in terms of the projected spin operators in Eqs.~\eqref{hpa1} and~\eqref{hpa2} as
\begin{equation}\label{ansi-ha1}
\begin{split}
H_{an}=&  -K\sum_{i}\left(S-a^\dagger_ia_i\right)^2-K\sum_j\left(S-b^\dagger_jb_j\right)^2 \\
=&-2NKS^2+\sum_i \left( 2KSa_i^\dagger a_i- Ka_i^\dagger a_i a_i^\dagger a_i \right)
 \\
 &+ \sum_i \left( 2KSb_j^\dagger b_j- Kb_j^\dagger b_j b_j^\dagger b_j \right)\\
 \approx & -2KNS^2+2KS\left( \sum_i a^\dagger_ia_i +\sum_j b_j^\dagger b_j    \right)
\end{split}.
\end{equation}
In the above, the fourth order terms $a_i^\dagger a_i a_i^\dagger a_i$ and $b_j^\dagger b_j b_j^\dagger b_j$ are also neglected. 
Based on the expressions in Eqs.~\eqref{reciprocal-cange1} and~\eqref{reciprocal-cange2}, the magnetic anisotropy term can be  reformulated  in the momentum space representation by
\begin{equation}\label{ansi-ha2}
H_{an}=-2KNS^2+2KS\sum_{\boldsymbol{k}} \left( a^\dagger_ka_k+ b^\dagger_kb_k \right).
\end{equation}

Eventually, the total Heisenberg Hamiltonian is written in the combined terms of intralayer (Eq.~\eqref{hpa4}) and interlayer (Eq.~\eqref{interlayer-ha-recp}) exchange coupling as well as magnetic anisotropy (Eq.~\eqref{ansi-ha2}) by 
     \begin{equation}\label{fianl-ham}\small{
     \begin{split}
         H(\boldsymbol{k})=&-\left[2K + Z_{ou}\left( J_1+J_2\right) +2Z_{in}J_{in}  \right]  NS^2 \\
         &+\sum_{k}S\left\{ 2K+Z_{ou}\left( J_1+J_2\right) +2Z_{in}J_{in}[1-\gamma_{in}(\boldsymbol{k})]   \right\} 
  \left(a_k^\dagger a_k+b_k^\dagger b_k \right)   \\
         &-\sum_kZ_{ou}S\{\left[J_1\gamma_1(\boldsymbol{k})+J_2\gamma_2(\boldsymbol{k}) \right] a_k^\dagger b_k  +h.c.    \}\\
     =&E_0+\sum_k\left[A_k\left(a_k^\dagger a_k+b_k^\dagger b_k \right) + \left(  B_k a_k^\dagger b_k+ h.c. \right)   \right]    
        \end{split}
        }.
\end{equation}

Here $E_0$, $A_k$ and $B_k$ are respectively rewritten by
\begin{equation}\label{fianl-ham-suppe}
\begin{cases}
E_0=-\left[2K + Z_{ou}\left( J_1+J_2\right) +2Z_{in}J_{in}  \right]NS^2   \\
A_k= S\left\{ 2K+Z_{ou}\left( J_1+J_2\right) +2Z_{in}J_{in}[1-\gamma_{in}(\boldsymbol{k})]   \right\}  \\
B_k=-Z_{ou}S\left[J_1\gamma_1(\boldsymbol{k})+J_2\gamma_2(\boldsymbol{k}) \right]
    \end{cases}
\end{equation}
Notably, the first term $E_0$ in Eq.~\eqref{fianl-ham} is the energy for the ground state.  Such term can be neglected in the  Hamiltonian because it makes no contribution to the spin wave dispersion.  
To obtain the matrix form of $k$-dependent Hamiltonian,   a set of  basis vectors and the corresponding Hermitian conjugate of the Boson operators are respectively defined by
\begin{equation}\label{bv1}
\Psi_k=(a_k,b_k)^T,\space \Psi_k^\dagger=(a^\dagger_k,b^\dagger_k) 
    \end{equation}
Neglecting the constant term, the Hamiltonian in  Eq.~\eqref{fianl-ham} can be written in terms of the quadratic form by $H(\boldsymbol{k})=\sum\limits_{\boldsymbol{k}}\Psi_k^\dagger \mathcal{M}(\boldsymbol{k}) \Psi_k$.  Based on Eqs.~\eqref{fianl-ham} and~\eqref{fianl-ham-suppe}, the dynamical matrix $\mathcal{M}(\boldsymbol{k})$ is expressed by
\begin{equation}\label{m-matrix}
\mathcal{M}(\boldsymbol{k})=\left[
\begin{array}{cc}
A_k & B_k  \\
B^*_k & A_k 
\end{array}
\right]
    \end{equation}
After diagonalization,  the obtained eigenvalues of the spin-wave dispersion are expressed by
\begin{equation}\label{eigenv}
\begin{aligned}
  \hbar\omega^\pm(\boldsymbol{k})=&A_k\pm \left|B_k \right|   \\
 =&S\left\{ 2K+Z_{ou}\left( J_1+J_2\right) +2Z_{in}J_{in}[1-\gamma_{in}(\boldsymbol{k})]   \right\} \\
 &\pm Z_{ou}S\left| J_1\gamma_1(\boldsymbol{k})+J_2\gamma_2(\boldsymbol{k})    \right|  
\end{aligned}
    \end{equation}

\subsection{Topological properties}
Similar to fermionic electrons, the bosonic magnon can also have topological properties~\cite{onose2010observation}
The topology of the magnon band is  evaluated by the Berry curvature~\cite{berry1984quantal}, which   determines the transverse motion of magnon wave packets under external fields hence the transport phenomena~\cite{qi2008topological,katsura2010theory,matsumoto2011theoretical}. Similarly,  the Berry curvature  for the $n$-th magnon band is written in term of  the curl of the
Berry connection~\cite{Berry-prm}
\begin{equation}\label{omega_n-berry}
\begin{aligned}
\Omega_n (\boldsymbol{k})  =\boldsymbol{\nabla}_{\mathbf{k}}\langle u_n(\boldsymbol{k}) |i \boldsymbol{\nabla}_{\boldsymbol{k}} |   u_n(\boldsymbol{k}) \rangle
\end{aligned},
    \end{equation}
where $u_n(\boldsymbol{k})$ is the corresponding Bloch state. For a 2D system, the anomalous transport properties are determined by the $xy$ component of the  Berry curvature, which is expressed by
\begin{equation}\label{omega_2-berry}
\begin{aligned}
\Omega_{xy}^{(n)}(\boldsymbol{k}) =
-2\,\mathrm{Im}
\left\langle
\frac{\partial u_n (\boldsymbol{k}) }{\partial k_x}
\Bigg|
\frac{\partial u_n (\boldsymbol{k})  }{\partial k_y}
\right\rangle
\end{aligned}.
    \end{equation}
Accordingly, the Chern number is evaluated in terms of the Berry curvature by
\begin{equation}\label{omega_2-berry}
\begin{aligned}
C_n=\frac{1}{2\pi i} \oint \Omega_{xy}^{(n)}(\boldsymbol{k}) d^2k
\end{aligned}.
    \end{equation}
For a two band model, the Berry curvatures of the two bands are canceled with each other  at an identical $K-$point ($\mathbf{K}=\boldsymbol{k}$), $i.e.$ $\Omega_{-}(\boldsymbol{k})+\Omega_{+}(\boldsymbol{k})=0$. where $\Omega_{\pm}$ in Eq.~\eqref{m-matrix} mark the acoustic and optical branches, respectively.  
In this regard, it is sufficient to calculate only the Berry curvatures  for the acoustic branch ($\Omega_{-}(\boldsymbol{k})$) as well as  Chern number.
Notably, for  visualization and regularization in numerical calculations, we introduce a small numerical mass term ($\Delta m=0.01$ meV) to lift the Dirac degeneracy, of which the value is much smaller than those of the exchange coupling and   magnetic anisotropy energy.

To further verify the topological properties,  we also calculate the edge states by constructing a  zigzag-edged ribbon with 80 unit cells along $y$  direction.  The edge states are obtained by means of  numerical diagonalization of the Hamiltonian matrix in real space~\cite{owerre2016first,mook2014edge,zhang2013topological}.

\section{Results and discussions}
\subsection{Magnetic properties and electronic structures}

The structure is relaxed in the FM spin configuration. In the  optimized structure, the lattice constant ($a$) is 2.825 \AA \space, while the distance ($d$) between the top and bottom layers of Fe is 2.017 \AA .
It is noticed that such structures are in good agreement with Ref.~\cite{YUE2017164}, where $a=$2.84 \AA \space and $d=$ 2.00 \AA. In addition, we also provide the bond angle information, as shown in Fig.~\ref{fig-1} (b). In detail, the bond angle of Fe(T)-C-Fe(B) (connecting the top and bottom layers of Fe via the ligand C) is $\alpha=85.1^\circ$,  while that of Fe(T)-C-Fe(T)  (connecting two neighbour Fe atoms in the top layer via the ligand C) is $\beta=94.9^\circ$. Obviously, the angles $\alpha$ and $\beta$ are supplementary due to inversion symmetry.

Based on the relaxed crystal structure, the ground spin configuration and the exchange coupling parameters are determined by comparing the energies of different spin configurations for Fe$_2$C. 
We have taken into account the NN intralayer exchange coupling, the NN and NNN interlayer exchange couplings between the top and bottom layers of Fe, which is  sufficient  to describe the Heisenberg Hamiltonian for magnetic MXene monolayers~\cite{LOU2022169959,limbu2025magnetic,agapov2024mxene}.  
To involve such exchange couplings, the $2\times2\times1$ super cell is  constructed, showing in Fig.~\ref{fig-1} (a). 
The considered spin configurations are shown in Fig.~\ref{fig-1} (c)--(f), including the ferromagnetic (FM) and three anti-ferromagnetic (AFM) phases. 
In AFM-A configuration, the NN intralayer exchange coupling is FM while the NN (NNN) interlayer exchange coupling is AFM (FM). By contrast, the AFM-C phase is in the configuration that NN intralayer  and NNN interlayer exchange couplings are both AFM while the NN interlayer exchange coupling is FM. Moreover, in AFM-G spin configuration, the NN intralayer and  NN interlayer exchange couplings are both AFM while the NNN interlayer exchange coupling is FM. Therefore, the exchange coupling parameters J$_{in}$, J$_{1}$ and J$_{2}$ are evaluated  based on the isotropic Heisenberg Hamiltonian
\begin{equation}\label{hei-ham-ani}
   H=-J_{in} \sum\limits_{i\neq j}\mathbf{S}_i\cdot\mathbf{S}_j -J_1\sum\limits_{k\neq l} \mathbf{S}_k\cdot\mathbf{S}_l-J_2\sum\limits_{m\neq n} \mathbf{S}_m\cdot\mathbf{S}_n,
\end{equation}
where $(i,j)$, $(k,l)$ and $(m,n)$ represents the NN intralayer, NN interlayer and NNN interlayer exchange coupling, respectively. 
Mapping DFT results into Eq.~\eqref{hei-ham-ani}, the energy differences between the AFM and FM  spin configurations are expressed by
\begin{equation}\label{hei-hama1}
\begin{cases}
\Delta_{A}=E_{AFM-A}-  E_{FM}= 6(J_{1}+J_2)S^2   \\
\Delta_{C}=E_{AFM-C}-  E_{FM}= 2(4 J_{in}+ J_1+3J_2)^2  \\
\Delta_{G}=E_{AFM-G}-  E_{FM}= 4(2 J_{in}+J_1)S^2      \\
\end{cases}.
\end{equation}
By solving the above equations, the exchange coupling parameters are expressed by the energy differences as
\begin{equation}\label{hei-exchange}
\begin{cases}
 J_{in}=\frac{1}{16S^2}\left( \Delta_C + \Delta_G - \Delta_A  \right)   \\
J_1 = \frac{1}{8S^2}\left( \Delta_A + \Delta_G - \Delta_C  \right)  \\
J_2 = \frac{1}{24S^2}\left( \Delta_A + 3\Delta_C -3 \Delta_G  \right)     \\
\end{cases}.
\end{equation}
The energy difference between  AFM and FM spin configurations are
$\Delta_{A}=$149.6 meV, $\Delta_{C}=$216.9 meV  and  $\Delta_{G}=$241.9 meV per chemical formula, respectively. 
The magnetic moment is 1.94 $\mu_B$ per iron atom. The quantum spin number is assumed to be S=2 for MXene Fe$_2$C~\cite{LOU2022169959,LI2021126960}.
Substituting the energy difference and quantum spin number, the resultant exchange parameters are $J_{in}=$4.847 meV, $J_{1}=$5.425   meV, and $J_{2}=$0.766 meV, respectively. Notably, the obtained exchange coupling parameters can be well compared with those reported by Lou~\cite{LOU2022169959} \textit{et al.}\ ($J_{in}=$5.24 meV, $J_{1}=$5.64   meV, and $J_{2}=$0.36 meV).
Clearly, the NN intralayer and NN interlayer exchange couplings are approximately 6--7 times larger than the NNN interlayer exchange coupling. 
This hierarchy directly reflects the exponential distance dependence of superexchange interactions, which dominate the magnetic coupling in Fe$_2$C MXene 
Moreover, the corresponding    distances of Fe-Fe connection are  
2.825 \AA \space (NN intralayer),   2.594 \AA \space (NN interlayer), and  3.835 \AA \space (NNN interlayer), respectively.  
Moreover, exchange interactions decay exponentially with increasing interatomic separation.  Given these facts, the longer-range exchange couplings beyond NNN interlayer exchange coupling would be at least an order of magnitude weaker than $J_2$, and thus are negligible for describing the magnetic ground state and low-energy excitations.
This suggests that the considered exchange couplings are sufficient to approach the Heisenberg Hamiltonian for MXene Fe$_2$C. 
In addition, the positive sign  indicates that such three exchange couplings are  FM. So, the following discussions are  based on the FM spin configuration.

The MAE is evaluated based on the energy difference between in-plane and out-of-plane magnetization as
\begin{equation}\label{mae}
  MAE=E_{[100]}-  E_{[001]},
\end{equation}
where $E_{[100]/[001]}$ denotes the energy for magnetization parallel to [100]/[001] direction. 
The obtained MAE for Fe$_2$C is -0.107 meV/f.u., indicating an in-plane magnetization.  
Such MAE value can be roughly comparable with the resultant MAE of 0.060 meV/f.u. reported by Lou~\cite{LOU2022169959} et al.
Furthermore, the contribution (parameter $K$ in Eq.~\eqref{ha1}) of each Fe is -0.054 meV.

\subsection{Magnon dispersion and topological properties }

\begin{figure*}
    \centering
    \begin{tabular}{cc}
         \includegraphics[width=0.35\linewidth]{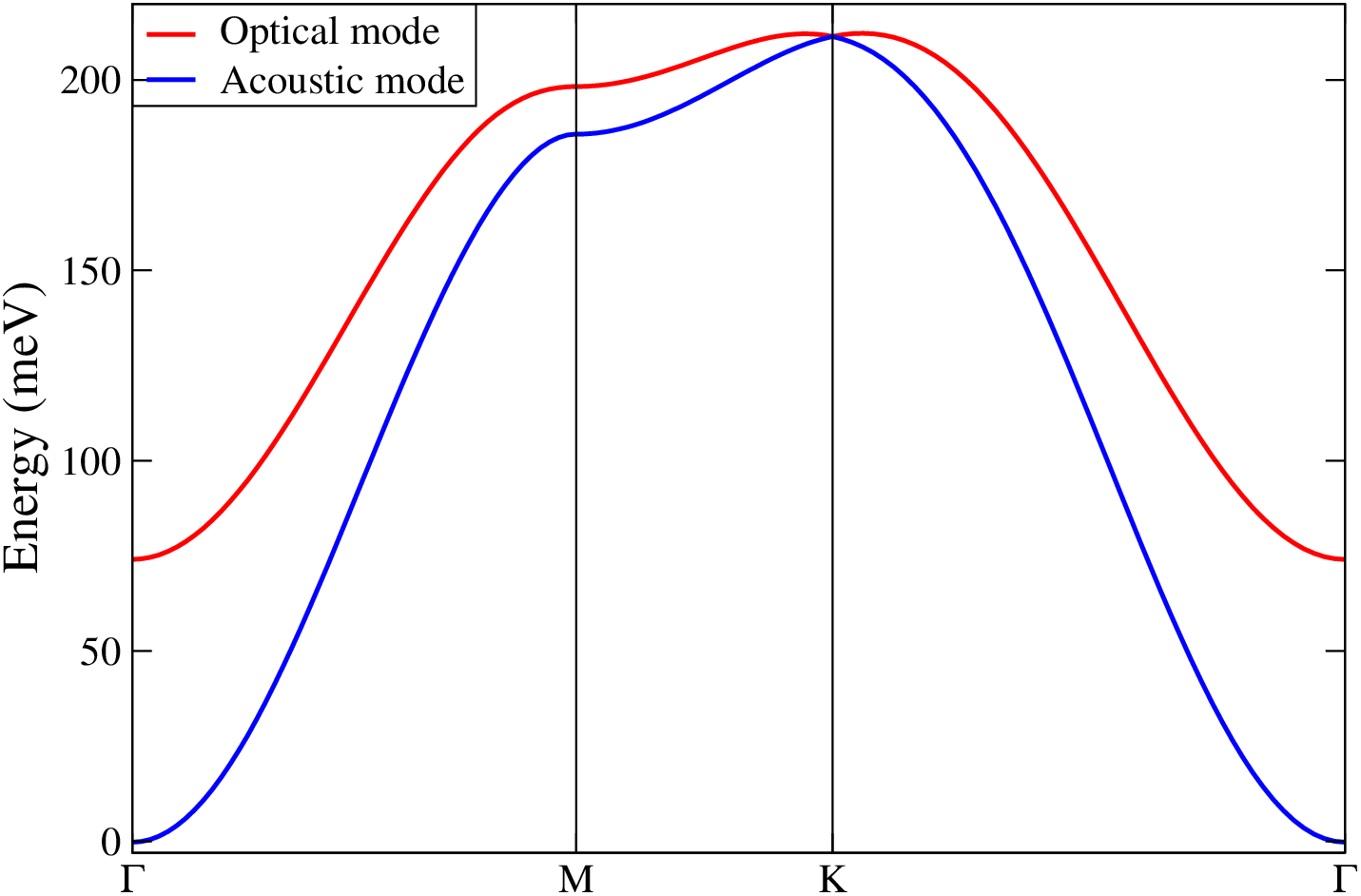} &   \includegraphics[width=0.35\linewidth]{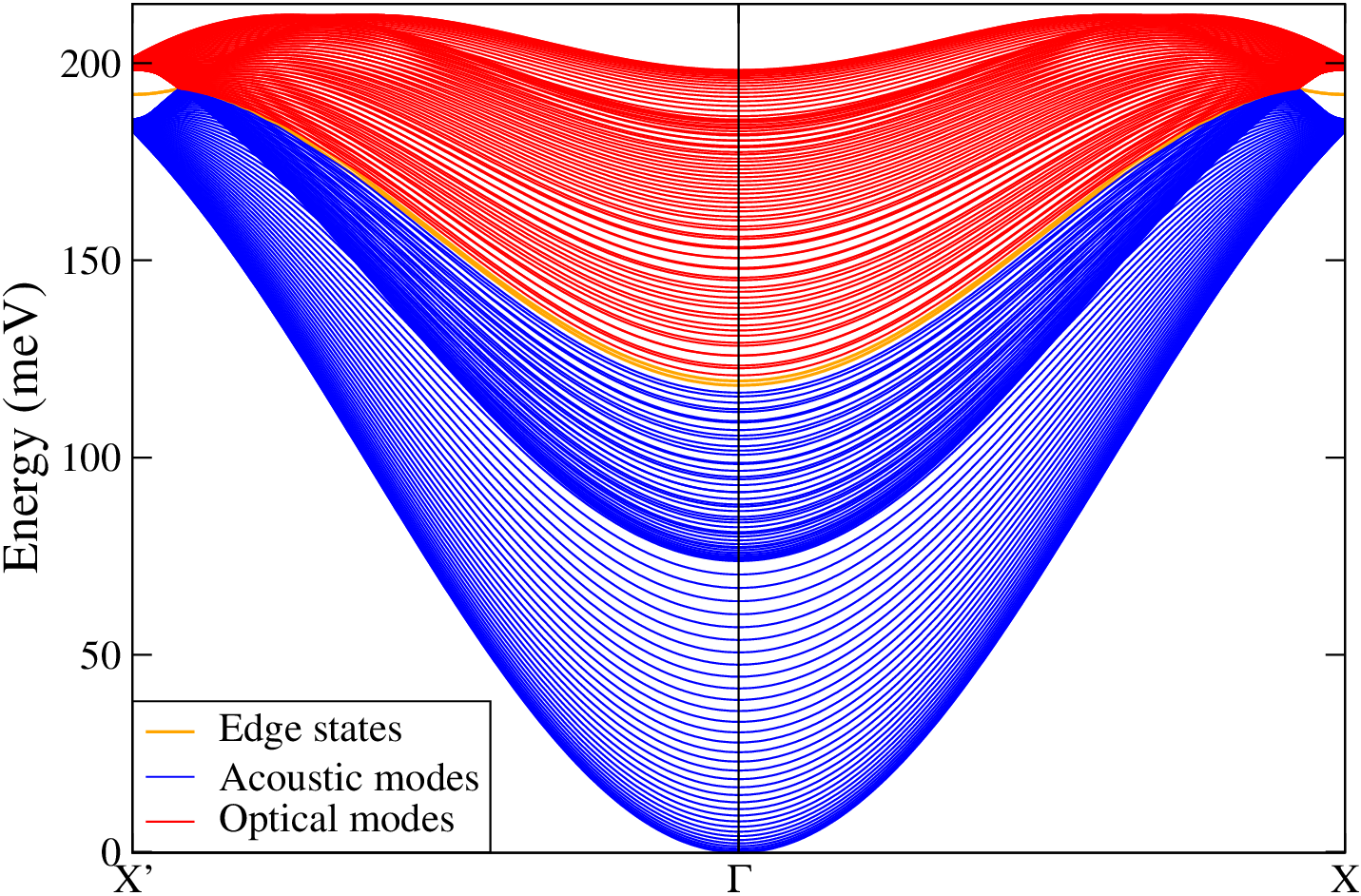}   \\ 
         (a) &  (b) \\
       \multicolumn{2}{c}{  \includegraphics[width=0.31\linewidth]{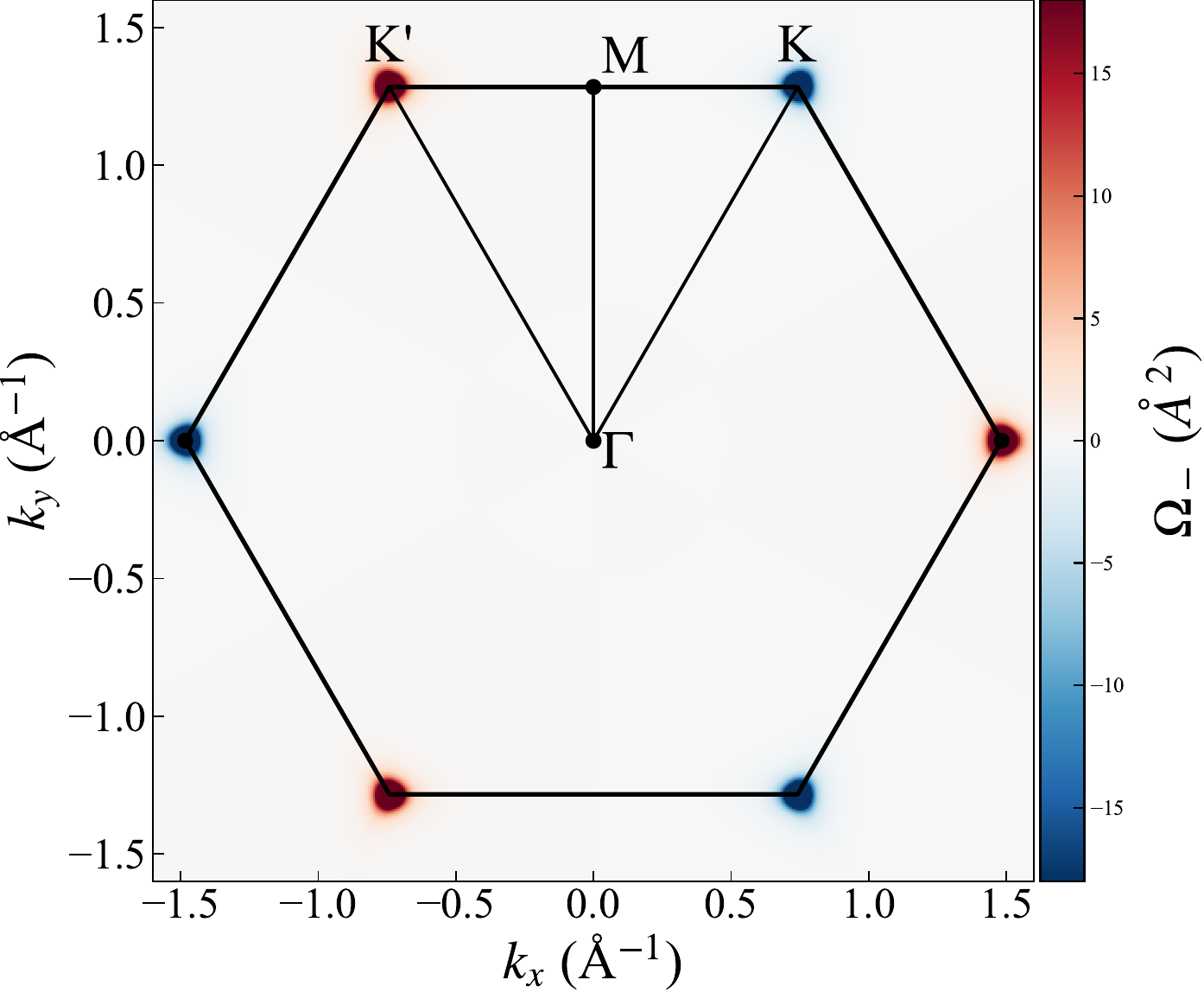} }\\ 
        \multicolumn{2}{c}{(c)}
    \end{tabular}
    \caption{(a) The magnon dispersion curves for MXene Fe$_2$C. The blue and red curves represent the  acoustic and optical modes, respectively. (b)  The magnon bulk bands  and  magnon edge modes using a zigzag-edged ribbon within 80 unit cells along $y$-direction. 
    (c) The Berry curvature of the acoustic  magnon band in the hexagonal Brillouin zone. Sharp peaks appear at the $K$ (0.5, 0, 0) and $K'$ ($-0.5$, 0, 0) corners, carrying opposite signs and corresponding to the Dirac degeneracy at 211.4 meV.  }
    \label{fig-sw1}
\end{figure*}

Showing in Fig.~\ref{fig-sw1} (a),  the magnon dispersion can be obtained by substituting the exchange coupling parameters ($J_{in}$, $J_{1}$ and  $J_{2}$) and magnetic anisotropy energy ($K=\frac{1}{2}MAE$) into Eq.~\eqref{eigenv}. 
It should be noticed that the magnitude of  the magnetic anisotropy term $K$ is much smaller than that of the exchange coupling. So, we treat the   magnetic anisotropy term as a small perturbation, leading to only an energy shifting in the magnon bands.
Different from degenerated bands in conventional FM materials, we can observe both  the acoustic and optical magnon bands in MXene Fe$_2$C, which is the reflection of   off‑diagonal term in Eq.~\eqref{eigenv} derived by interlayer exchange coupling. 
‌
Although separated by the local  gaps of  74.3  and 18.6 meV at  $\Gamma$   and $M$  points, there is not a global gap between the  acoustic and optical  magnon  branches.  
In particular, such acoustic and optical magnon  bands crosses with each other at the $K$ point  211.4 meV, forming a magnon Dirac point. Topological properties have also been predicted in the other 2D monolayers~\cite{brehm2024topological,schneeloch2022gapless,eto2026fate}.

To confirm the topological properties,  
the Berry curvature for the acoustic branch is  shown in Fig.~\ref{fig-sw1} (c) at the 211.4 meV, which corresponds exactly to the energy level for the band crossing point  at the $K$ point.  
The Berry curvature vanishes  almost throughout the whole Brillouin zone  except for the corners $K$ and $K'$. 
Around $K$ and $K'$ points, the  Berry curvatures display  a pronounced sharp peak with  an identical magnitude but the opposite signs due to inversion symmetry.
Furthermore, the total Chern number is $zero$ over the whole Brillouin zone, while the valley-resolved Chern numbers are   $+\frac{1}{2}$  and $-\frac{1}{2}$ at the  $K$ and $K'$ valleys, respectively.
As depicted in Fig.~\ref{fig-sw1}(b), 
we can observe two topological non-trivial edge states    in the  boundary spectrum.  
Such two valley-momentum locked edge states are protected by 
the valley Chern numbers, leading to the opposite transverse 
velocities for magnon bands in the $K$ and $K'$ valleys under an in-plane temperature gradient. 
In summary, the crossing point between acoustic and optical magnon bands is a  topological non-trivial Dirac point with the valley Chern numbers of $\pm\frac{1}{2}$ at $K$ and $K'$ points.  The opposite transverse velocities in the two valleys under an in-plane temperature gradient  give rise to the magnon valley Hall effect~\cite{faizee2026magnon,bai2026ferroelectrics,xing2022valley}. These results demonstrate that Fe$_2$C can be designed for future high-efficiency  valleytronic devices.
Notably, the  degeneracy at K point and  magnon topology  have also been discovered in the model study for   two-dimensional ferromagnetic  AB-stacking bilayers of  honeycomb lattice~\cite{ortmanns2021magnon}. Similarly,   the Fe atoms in the structure of  MXene Fe$_2$C system  can also be regarded as AB-stacking on top and bottom of C layers, where Fe of each sub-lattice is in the trigonal  lattice. This  AB-stacking will induce interlayer exchange coupling as well as  degeneracy at K point in magnon bands~\cite{ortmanns2021magnon}.

To better understand the topological properties, we will construct an effective model around $K$ point under low energy limit.
To begin with  symmetry, the MXene Fe$_2$C crystallizes in  the space group of P$\overline{3}$m1 (No. 164), including the  three‑fold rotation $C_3$ about $c$-axis and spatial inversion $P$. 
Under $C_3$ rotation, the high-symmetry $K$  ($\frac{2\pi}{3a}$, $\frac{2\sqrt{3}\pi}{3a}$, 0) point in the Brillouin zone   is  invariant as 
\begin{equation}\label{c3k}
C_3\mathbf{K} = \left(-\frac{4\pi}{3a},\; 0\right) = \mathbf{K} + \mathbf{G},
\end{equation}
where  $\mathbf{G}=(\frac{2\pi}{a}, \frac{2\sqrt{3}\pi}{3a}, 0)$ is a primitive reciprocal lattice vector. 
Due to the destructive interference derived by the $C_3$ rotational symmetry,  the interlayer structure factors $\gamma_1(\boldsymbol{k})$ and $\gamma_2(\boldsymbol{k})$ in Eq.~\eqref{structure-factor} are both $zero$  at  $K$ point, which does not rely on the specific parameters. Consequently,  the interlayer exchange coupling term $B(k)$  in Eq.~\eqref{eigenv}   also vanishes  at $K$ point, $i.e.$ $B(\mathbf{K})=0$.
Applying inversion symmetry $P$, the $K$ point will become $\mathbf{K}'=-\mathbf{K}$. Considering $B(-\boldsymbol{k})=B^*(\boldsymbol{k})$, we can  easily  obtain  the condition
\begin{equation}\label{bkzero}
B(\mathbf{K}')=B(\mathbf{K})=0.
\end{equation}
Overall, the Dirac degeneration of the magnon for Fe$_2$C at $K$ (and $K'$) point is topologically non-trivial, which is protected by the  rotational symmetry $C_3$. 
Moreover, for   an arbitrary wavevector $\boldsymbol{k}$ around $K$ point,  the offset displacement wavevector is determined by   $\boldsymbol{q}=\boldsymbol{k}-\boldsymbol{K}=(q_x,q_y,0)$. 
So, the interlayer exchange coupling term $B(k)$ can be expanded  in a Taylor series by
\begin{equation}\label{talyer-bk}
\begin{aligned}
B(\boldsymbol{k}) = &B(\boldsymbol{K}+\boldsymbol{q})   \\
  \approx &  B(\boldsymbol{K}) +  \left. \left(
\frac{\partial B}{\partial  q_x}q_x
+ \frac{\partial B}{\partial q_y}q_y
\right)  \right|_{\boldsymbol{q}_0}  \\
=& \alpha q_x +\beta q_y+\mathcal{O}(q^2)
\end{aligned},
\end{equation}
where $\boldsymbol{q}_0=(0,0,0)$, representing the $K$ point.  The above equation makes use of the vanishing relation in Eq.~\eqref{bkzero}.  
 The  coefficients $\alpha$ and $\beta$ are explicitly expressed using the partial derivatives for the interlayer exchange coupling structure factors $\gamma_{1/2}$ (in Eq.~\eqref{structure-factor})  by the relation 
\begin{equation}\label{taylerbk-sup}
\begin{aligned}
\alpha &\equiv \left.\frac{\partial B}{\partial  q_x}\right|_{\mathbf{q}_0}
       = -Z_{\text{ou}}S \left.   \left( J_1\frac{\partial\gamma_1}{\partial q_x}
                            + J_2\frac{\partial\gamma_2}{\partial q_x}  \right) \right|_{\boldsymbol{q}_0   }\\[4pt]
      &= \sqrt{3}Z_{\text{ou}} S a\left( \frac{J_1}{6}\,e^{i\pi/3} + \frac{J_2}{3}\,e^{-i2\pi/3} \right), \\[8pt]
\end{aligned}
\end{equation}

\begin{equation}\label{taylerbk-sup-2}
\begin{aligned}
\beta  &\equiv \left.\frac{\partial B}{\partial q_y}\right|_{\mathbf{q}_0}
       = -Z_{\text{ou}}S  \left. \left( J_1\frac{\partial\gamma_1}{\partial q_y}
                            + J_2\frac{\partial\gamma_2}{\partial q_y} \right) \right|_{\boldsymbol{q}_0}  \\[4pt]
      &= -\sqrt{3}i Z_{\text{ou}} S a\left[ \frac{J_1}{18}\,e^{i\pi/3}
           + \left( -\frac{J_1}{9} + \frac{J_2}{3} \right) e^{-i2\pi/3} \right].
\end{aligned}
\end{equation}
Similarly, the intralayer exchange structure factor   $\gamma_{in}(\boldsymbol{k})$  in Eq.~\eqref{structure-factor} can also be expanded  in a Taylor series, yielding
\begin{equation}\label{talyer-gammain}
\begin{aligned}
\gamma_{\text{in}}(\boldsymbol{k}) &= \gamma_{\text{in}}(\boldsymbol{K}+\boldsymbol{q}) \\
&\approx \gamma_{\text{in}}(\boldsymbol{K}) 
+\textstyle \frac{1}{2} \left.  \left( \frac{\partial^2\gamma_{\text{in}}}{\partial q_x^2}q_x^2 
+ 2\frac{\partial^2\gamma_{\text{in}}}{\partial q_x\partial q_y}q_x q_y 
+ \frac{\partial^2\gamma_{\text{in}}}{\partial q_y^2}q_y^2 \right)
 \right| _{\!\mathbf{q}_0} \\
&= \gamma_{\text{in}}(\boldsymbol{K}) + \kappa \,q^2 + \mathcal{O}(q^3)
\end{aligned}.
\end{equation}
where $q^2=|\mathbf{q}|^2=q_x^2+q_y^2$.  Correspondingly, the intralayer exchange coupling term $A (\boldsymbol{k})$ can be reformulated by
\begin{equation}\label{talyer-akexpan}
\small
\begin{aligned}
A(\boldsymbol{k}) &= A(\boldsymbol{K}+\boldsymbol{q}) \\
&\approx A(\boldsymbol{K}) 
+ {\frac{1}{2} \left. \left( \frac{\partial^2A}{\partial q_x^2}q_x^2 
+ 2\frac{\partial^2A}{\partial q_x\partial q_y}q_x q_y 
+ \frac{\partial^2A}{\partial q_y^2}q_y^2 \right) \right|_{\boldsymbol{q}_0}} \\
&= A(\boldsymbol{K}) 
- {6 S J_{\text{in}} \left. \left( \frac{\partial^2\gamma_{\text{in}}}{\partial q_x^2}q_x^2 
+ 2\frac{\partial^2\gamma_{\text{in}}}{\partial q_x\partial q_y}q_x q_y 
+ \frac{\partial^2\gamma_{\text{in}}}{\partial q_y^2}q_y^2 \right) \right|_{\boldsymbol{q}_0}} \\
&= A(\boldsymbol{K}) - \xi\,q^2 + \mathcal{O}(q^3).
\end{aligned}
\normalsize
\end{equation}
where $\xi=\frac{3}{2}SJ_{in}a^2$ and $A(\boldsymbol{K}) = S[2K + 3(J_1+J_2) + 18J_{\text{in}}]$.

In this part, we will analyze both the interlayer and intralayer exchange coupling terms   from the perspective of symmetry. 
At the $K$ ($\mathbf{K}=\left(\frac{2\pi}{3a},\frac{2\pi}{a\sqrt{3}},0\right)$)  point, the diagonal term $A(\boldsymbol{k})$ exhibits vanishing first‑order derivatives,  since $\gamma_{\text{in}}$ attains an extremum. 
As mentioned in the last paragraph,  the first‑order derivatives of the interlayer exchange coupling term B($\boldsymbol{k}$) (or $\gamma_{1/2}(\boldsymbol{k})$)  are nonzero  at the same point. 
Such  distinct  behaviors can be explained from the perspective of symmetry.    The $K$ point 
is invariant under the  three‑fold rotation $C_3$, so its little 
group is $C_{3v}$ (a subgroup of the full point group $D_{3d}$ of the 
$P\overline{3}m1$ space group).
A small momentum displacement $\boldsymbol{q}=(q_x,q_y)$ in the vicinity of the $K$ point transforms
under the 2D irreducible representation $E$ of $C_{3v}$.
The intralayer structure factor $\gamma_{\text{in}}(\boldsymbol{k})$ is a real scalar  function that is invariant under all symmetry operations of the crystal. Therefore it belongs to the fully symmetric representation $A_1$.
At a high‑symmetry point, the gradient of an \(A_1\) scalar with respect to $E$‑transforming coordinates must vanish. A nonzero gradient would produce an $E$ component, contradicting the function’s \(A_1\) symmetry.   Consequently,  the Taylor expansion of $A(\boldsymbol{k})$ ($\gamma_{\text{in}}$) starts from the quadratic term.
In contrast, the interlayer structure factors $\gamma_{1/2}(\boldsymbol{k})$  
 together form a 
basis  of the   2D irreducible representation $E$ of $C_{3v}$ at the $K$ point.  
The real and imaginary components of an $E$-symmetric function can couple linearly to the  momentum components $q_x, q_y$ (which also transform as $E$) to form an $A_1$ 
invariant, where the linear coupling is responsible for the   Dirac cone in the magnon
dispersion.
Therefore, the first‑order derivatives of $B(\boldsymbol{k})$  ($\gamma_{1/2} (\boldsymbol{k})$ )  are symmetry‑allowed and generally 
non‑zero.  Furthermore, the second‑order derivatives of $B(\boldsymbol{k})$  ($\gamma_{1/2} (\boldsymbol{k})$ ) can be neglected 
since such high order corrections  do not alter the linear band crossing or the topological invariants.
To summarize, at $K$ ($\mathbf{K}=\left(\frac{2\pi}{3a},\frac{2\pi}{a\sqrt{3}},0\right)$)  point,  the   interlayer and intralayer exchange coupling terms ($B(\boldsymbol{k})$  and $A(\boldsymbol{k})$) are  expanded by the
Taylor series  up to   linear and  quadratic orders, respectively.  
Such distinct behaviors  are driven by  the different symmetry representations of $\gamma_{\text{in}}$ and  $\gamma_{1/2}$, leading to the Dirac cone and   topological nontrivial   magnon bands in Fe$_2$C.

The diagonal intralayer exchange coupling term $A(\boldsymbol{k})$  only adds an overall energy shift and a trivial quadratic warping to both   acoustic and optical  bands. This means the contribution from  $A(\boldsymbol{k})$ can be regarded as a constant value (211.4 meV)  in a small vicinity of the \(\boldsymbol{K}\) point when deriving the effective Hamiltonian. Thus, the  effective Hamiltonian can be rewritten in terms of the Pauli matrix $\boldsymbol{\sigma}=(\sigma_x, \sigma_y)$ by 
\begin{equation}\label{eff-hamd}
\begin{aligned}
  \mathcal{H}_\text{eff}(\boldsymbol{q})&=\begin{bmatrix}
 0 &  B(\boldsymbol{k})  \\
   B^*(\boldsymbol{k}) &  0
\end{bmatrix}=[\Re B(\boldsymbol{q})]\sigma_x-[\Im B(\boldsymbol{q}) ]\sigma_y\\
  &=\boldsymbol{d(q)}\cdot \boldsymbol{\sigma}
\end{aligned}.
\end{equation}
Here $\Re$ and $\Im$ are the real and imaginary  component functions, respectively.  Due to $d_z(\boldsymbol{q})\equiv0$, the $z$  component  is suppressed, which reflects the chiral symmetry
$\sigma_z\mathcal{H}_{\text{eff}}\sigma_z = -\mathcal{H}_{\text{eff}}$ and
the absence of any Dzyaloshinskii-Moriya interaction~\cite{owerre2016first,pershoguba2018dirac}.
We can define the pseudo-spin vector $\boldsymbol{d(q)}$ by
\begin{equation}\label{dmatrix}
\boldsymbol{d(q)}=\begin{bmatrix}
  d_x(\boldsymbol{q})  \\
  d_y(\boldsymbol{q})  \\  
\end{bmatrix}=\begin{bmatrix}
  \alpha_R  &\beta_R   \\
   -\alpha_I  &-\beta_I   \\
\end{bmatrix}\begin{bmatrix}
  q_x   \\
  q_y   \\
\end{bmatrix}=M\begin{bmatrix}
  q_x   \\
  q_y   \\
\end{bmatrix},
\end{equation}
 where the subscripts $I$ and $R$ mark the imaginary and real parts of $\alpha$ (or $\beta$), respectively.
Therefore,  the  effective Hamiltonian 
\begin{equation}\label{eff-ham-sigam}
\mathcal{H}_{\mathrm{eff}}(\mathbf{q}) = \begin{bmatrix}
0 & \alpha q_x + \beta q_y \\
\alpha^* q_x + \beta^* q_y & 0
\end{bmatrix}
= \mathbf{d}(\mathbf{q})\cdot\boldsymbol{\sigma}.
\end{equation}

To diagonalize $M$, we introduce a canonical orthogonal transformation  \(R\in\mathrm{O}(2)\), which defined by  $\boldsymbol{q}' = R\boldsymbol{q}$.
Then the  pseudo-spin vector can be rewritten by 
\begin{equation}\label{cano-trans}
     \boldsymbol{d(q')} = MR^\mathrm{T} \begin{pmatrix} q_x' \\ q_y' \end{pmatrix}.
\end{equation}
By choosing an appropriate  rotation angle, the matrix product \(MR^\mathrm{T}\) can be cast into a strictly diagonalized canonical form:
\begin{equation}\label{cano-trans-mrt}
M R^{T} = \begin{pmatrix} v_x & 0 \\ 0 & -v_y \end{pmatrix}, \qquad v_x,v_y > 0,
\end{equation}
Correspondingly, the effective Hamiltonian becomes the standard canonical anisotropic Dirac model format as
\begin{equation}\label{cano-tra-dq}
\mathcal{H}_\mathrm{eff}(\boldsymbol{q}') = v_x q_x' \sigma_x + v_y q_y' \sigma_y= \boldsymbol{ d({q}')}\cdot  \boldsymbol{\sigma}   
\end{equation}
where the new pseudo-spin vector is defined by $\boldsymbol{ d({q}')} = \big(v_x q_x', -v_y q_y'\big)$. The magnon dispersion  is subsequently expressed in the canonical energy spectrum as 
\begin{equation}\label{newomegapm}
    \omega_\pm(\boldsymbol{q}') = \pm\sqrt{v_x^2 q_x'^2 + v_y^2 q_y'^2}
\end{equation}

For a generic two‑band model, the  Berry curvature of the acoustic magnon band ($\omega_-$)  is evaluated by~\cite{berry1984quantal,qi2008topological,owerre2016first}
\begin{equation}
\Omega_{-}(\boldsymbol{q}') = -\frac{1}{2}\,
\frac{\boldsymbol{d}\cdot\bigl(\partial_{q_x'}\boldsymbol{d} \times \partial_{q_y'}\boldsymbol{d}\bigr)}
{|\boldsymbol{d}|^{3}}.
\label{berry-formula}
\end{equation}
Using $d_z=0$,  pseudo-spin vector is reformulated by $\boldsymbol{ d({q}')} = \left(v_x q_x', -v_y q_y', 0 \right)$ under the canonical  transformation.  Then the numerator in Eq.~\eqref{berry-formula} is evaluated  as  
$\boldsymbol{d}\cdot(\partial_{q_x'}\boldsymbol{d} \times \partial_{q_y'}\boldsymbol{d}) = -v_x v_y\,d_z = 0$ since $d_z=0$.    
Apparently, the Berry curvature is always $zero$ around $K$ point except for the
Dirac point, which is a singularity with $\left|q'_0\right|=0$.  
To approach the  Berry curvature at the Dirac point, we  introduce a small mass to the $z$ component of pseudo vector  as  $\boldsymbol{ d({q}')}=\left(v_x q_x', -v_y q_y', \Delta m \right)$.
Eventually, the regularized Berry curvature at the Dirac point is computed by 
\begin{equation}
\begin{aligned}
\lim_{\Delta m \to 0}  \Omega_{-}(\boldsymbol{q}') &= \lim_{\Delta m \to 0} \frac{v_x v_y\, \Delta m}
{2\,\bigl(v_x^2 q_x'^{2} + v_y^2 q_y'^{2} + \Delta m^{2}\bigr)^{3/2}}\\   
&=\pi  \mathrm{sgn}(\Delta m )\delta(\boldsymbol{q}')
\end{aligned}.
\label{berry-formula-regula}
\end{equation}
When  $\Delta m \to 0$,  the  Berry curvature behaves as a delta function. 
It varies smoothly  throughout the entire     Brillouin zone except for  a dramatic sharp peak at the Dirac point ($\boldsymbol{q}'=\boldsymbol{q}_0$), as shown in Fig.~\ref{fig-sw1} (c).  The sign of the peak is identical to that of $\Delta m$. As a result of inversion symmetry, the Berry curvature at the inequivalent valley is 
\begin{equation}
\Omega_{-}(\boldsymbol{K'}) =  - \Omega_{-}(\boldsymbol{K}).
\label{berry-formula-kprime}
\end{equation}
This is the reason for the opposite signs  of the Berry curvature at $K$ and $K'$ points in Fig.~\ref{fig-sw1} (c).

The Berry phase $\gamma_\text{Berry}$ can be obtained by integrating Eq.~\eqref{berry-formula-regula} over the full momentum space as
\begin{equation}
\gamma_\text{Berry}=\int_{\mathbb{R}^2} \Omega_{-}(\boldsymbol{q}')\,d^2q' =  \pi  \mathrm{sgn}(\Delta m ),
\label{berry-phase}
\end{equation}
where $\mathrm{sgn}$ is the sign function.  Clearly, the magnitude of the Berry phase is independent of that of  $\Delta m$, while the sign of the former is the same as the latter.   
More importantly, the magnon wave packet encircling the Dirac point acquires a Berry phase of $\pm\pi$, indicating a non-trivial winding number of $\pm1$ for the pseudo-spin vector in momentum space. The non-trivial winding number $\pm1$ can also be confirmed by the pair of chiral edge modes in the ribbon as shown in Fig.~\ref{fig-sw1} (b).  
Furthermore, the valley Chern number  can be obtained from the Berry phase around a single valley by
\begin{equation}
C_{K/K'}=\frac{\pm1}{2\pi}\int_{K/K'}\Omega_{-}(\boldsymbol{k})d^2k=\frac{\pm1}{2\pi}\gamma_\text{Berry}=\frac{\pm\mathrm{sgn}(\Delta m)}{2}.
\label{chern-numb}
\end{equation}
So, the total Chern number over the whole Brillouin zone is
\begin{equation}
C_{K}+C_{K'}=0,
\label{chern-numb-sum}
\end{equation}
Though with a vanishing total Chern number,  the acoustic mode for Fe$_2$C magnon has the  valley Chern numbers of $\pm\frac{1}{2}$.

\subsection{superexchange mechanism}

\begin{figure*}
\centering
\begin{tabular}{cc}
  \includegraphics[scale=0.38]{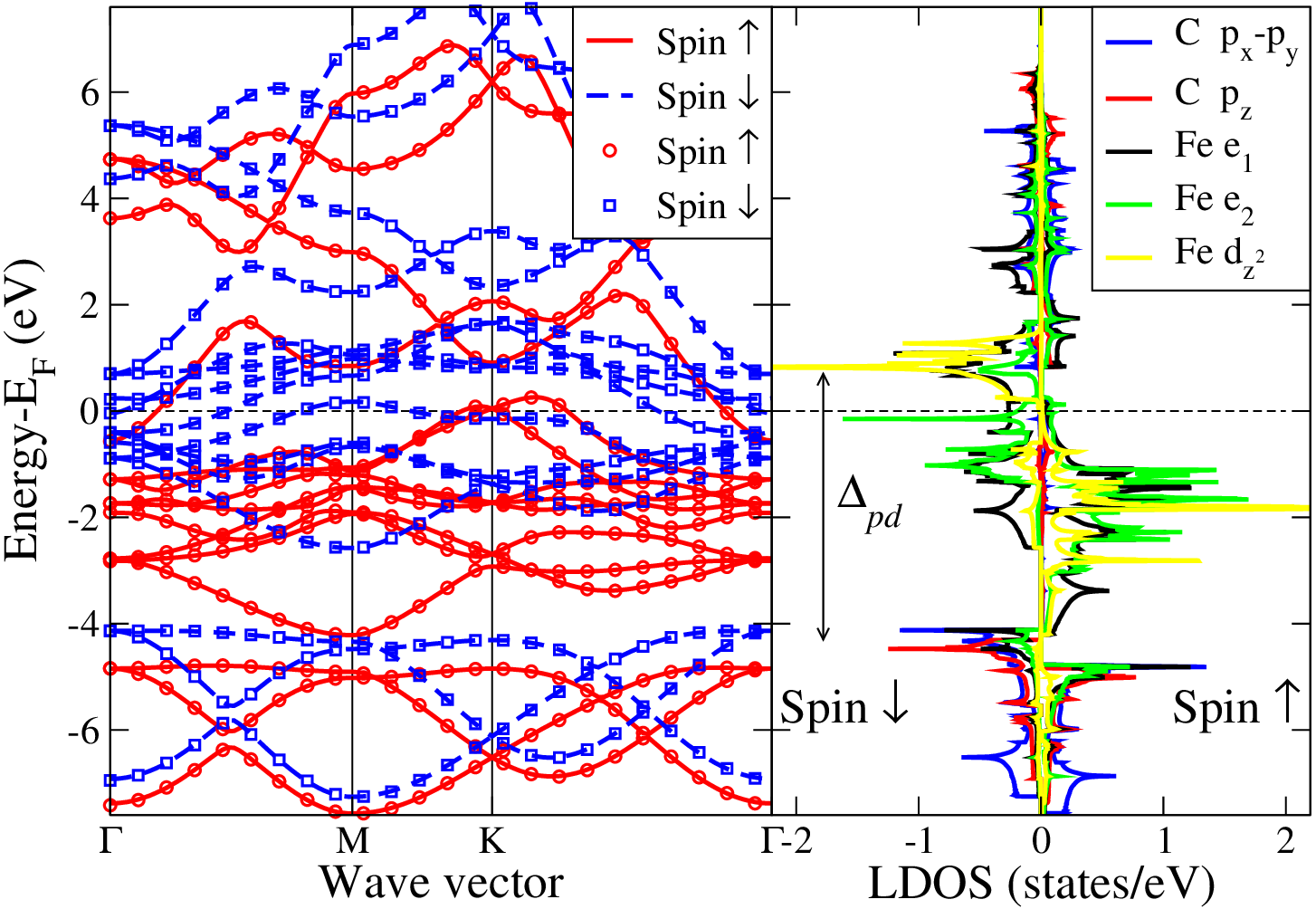}     & \includegraphics[scale=0.38]{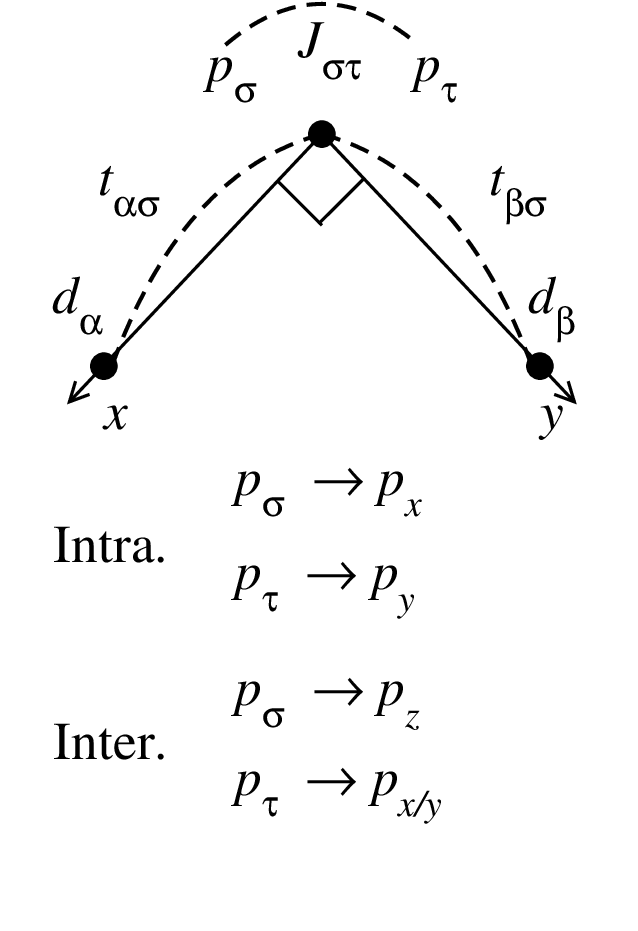}  \\
   (a)  & (b)
\end{tabular}
  \caption{(a) The band structures (left) for MXene Fe$_2$C together with the local density of states (LDOS)  (right).  
 In the band structures, the solid red and dashed blue curves denote the band structures of spin up and down channels by using DFT calculations. The red circle and blue square curves are the band structures of spin up and down channels based on Wannier interpolation. In LDOS, for Fe, $e_1$  and $e_2$ represent the doubly degenerated states   (d$_{xy}$-d$_{x^2-y^2}$) and e$_2$ (d$_{yz}$-d$_{xz}$) while  $d_{z^2}$   is a singlet state. For C, the p orbital is split into doubly degenerated $p_x-p_y$ and singlet $p_z$.  The black horizontal dashed line denotes the Fermi level. $\Delta_{pd}$ marks the charge transfer energy during the virtual hopping process. (b) The sketch for superexchange path in Fe$_2$C.  $t_{\alpha\sigma/\beta\tau}$  denotes the hopping between $d_{\alpha/\beta}$ orbital of Fe and $p_{\sigma/\tau}$ orbital of C.  $J_{\sigma\tau}$ is the Hund exchange integral between the orthogonal $p_\sigma$ and $p_\tau$ orbitals.
 }\label{fig2}
\end{figure*}

The interlayer exchange coupling hybridizes the magnon modes of the two magnetic sub-lattices in the bottom and top Fe layers of Fe$_2$C,  thereby lifting the degeneracy between the acoustic and optical branches. 
The exchange couplings (both intralayer and interlayer) directly govern the spin wave behavior and the topological properties.  In atomic scale,  the superexchange plays an essential role to determine the exchange couplings.  Thus, we will construct a model to figure out a crucial insight into the superexchange for Fe$_2$C,  providing  a theoretical basis for the manipulation of spin wave.

The microscopic origin of  superexchange can be clarified based on the kinetic exchange mechanism via $pd$-hoping between the magnetic and ligand atoms~\cite{li2025room,gubo-2023-super}.
So, the symmetry-adapted Wannier functions for Fe$_2$C are constructed based on the electronic structures obtained by using FPLO code~\cite{koepernik1999full,opahle1999full}. As shown in Fig.~\ref{fig2}, the band structure of Wannier function model can be well compared with that of DFT calculations.  
The mean least square deviation of Wannier function~\cite{zhang2018high} is evaluated by $\sum=\frac{100}{N} \sum\limits_{i\in occ.} \left( E^{DFT}_i-E_i^{WAN} \right)^2$,
where the $E^{DFT/WAN}_i$ denotes the eigenvalue for the $i-$th occupied band by means of direct DFT calculations and Wannier interpolation.  A scaling factor is applied for better readability. The mean least square deviation for Fe$_2$C is 4.9355$\times 10^{-3}$ eV$^2$, indicating the fine Wannier interpolation. In addition, the electronic bands crosses Fermi level in both channels, indicating a metallic ferromagnetic feature. It is reported~\cite{YUE2017164} that the  itinerant ferromagnetic order obeys the Stoner's  criterion.

To unravel the intrinsic mechanism governing superexchange, we have studied the   orbital occupation by using the local density of states  (LDOS) in Fig.~\ref{fig2} (a).
The local density of states is shown  to clarify the orbital occupation. In general, the MXene compounds belong to a 3m point group, corresponding to a $C_{3}$ crystallographic symmetry. Under  such C$_{3v}$ crystal field, the 3d orbitals of Fe  split into three distinct manifolds, namely the doubly degenerated states of e$_1$ (d$_{xy}$-d$_{x^2-y^2}$) and e$_2$ (d$_{yz}$-d$_{xz}$) together with the non-degenerated state of d$_{z^2}$.  In parallel, the doubly degenerated p$_x$-p$_y$ and the singlet p$_z$ states are observed in splitting of the 2p orbital of C. 
 Specifically,  the full set of 3d orbitals for Fe in the spin-up channel is nearly fully occupied, with only a minor fraction  of e$_1$ manifold remains unoccupied. 
 For the spin-down channel, the   d$_{z^2}$ orbital can be regarded as nearly completely unoccupied, leading to a contribution of about 1 $\mu_B$ to the magnetic moment.
 Furthermore, both e$_1$ and e$_2$ manifolds are partially filled, where the occupation of e$_2$ state is a bit more  than that of e$_1$ state. The  e$_1$ and e$_2$ manifolds make a total contribution of approximately 1 $\mu_B$ to the magnetic moment. Thus, the magnetic moment is about 2 $\mu_B$ in each Fe atom. 
In addition, the spin splitting of the 2p orbitals of C is much weaker than that of Fe.  
Obviously, the main part of 2p orbitals of C is away from Fermi level with an energy level in the range from -7 eV to -4 eV, while that of the 3d orbital of Fe crosses Fermi level with an energy in the range from -2.5 eV to 2 eV in the spin down channel and -4 eV to 0 eV in the spin up channel.  
There is a remarkable   gap between 3d orbital of Fe and 2p orbital of C in the spin down channel. By contrast, in the spin up channel, these   two orbitals almost cross with each other. Such dramatically different behaviors of LDOS in spin up and down channels can be ascribed to the strong exchange splitting of the d orbital of Fe. 
It should be noticed that such LDOS can be comparable with Refs.~\cite{YUE2017164,LOU2022169959}.

The underlying mechanism of superexchange  can be elucidated based on the Goodenough-Kanamori-Anderson (GKA) rules~\cite{Goodenough-1955,KANAMORI195987,Anderson-1950,Anderson-1959}. The Fe-C-Fe bond angles that connect the interlayer and intralayer exchange couplings  are respectively $\alpha$=85.1$^\circ$ and $\beta$=94.9$^\circ$.  According to GKA rules, such approximately 90$^\circ$ cation-anion-cation connections lead to weak ferromagnetic couplings. The superexchange can be  reasonably demonstrated by a simple Fe$_1$-C-Fe$_2$ model~\cite{qian2017ferromagnetism},  illustrating in Figure~\ref{fig2} (b). For each Fe\(_1\)–C–Fe\(_2\) path, a local Cartesian coordinate system is constructed by placing the bridging C atom at the origin, with Fe\(_1\) on the x-axis and Fe\(_2\) on the y-axis.
The indirect exchange interaction  is mediated by the virtual hopping of  two electrons from two occupied orthogonal p orbitals of the C atom  into the unoccupied d orbitals of two neighbouring Fe atoms. 
Crucially, this virtual hopping  is restricted to the spin-down channel, since the d orbitals of Fe in the spin-up channel are nearly fully occupied and thus Pauli-blocked  while those in the spin-down channel  remain partially unoccupied   as shown in Fig.~\ref{fig2} (a). On the other hand, under the C$_{3v}$ crystal field, the 3$d$ orbital of Fe can be classified by  three  non-equivalent irreducible representations (irreps), namely the non-degenerated irrep $A_1$ \{$d_{z^2}$\}  as well as doubly degenerated  $E$ irreps   \{d$_{xy}$, d$_{x^2-y^2}$\} and \{d$_{yz}$, d$_{xz}$\}. Similarly,  
the 2$p$ orbital of C will be split into the non-degenerated $A_1$  basis p$_z$   and doubly degenerated $E$ basis \{p$_x$, p$_y$\}.

In the following, we will derive  the exchange coupling based on perturbation theory following the spirit in Ref.~\cite{zhang-2022-gak}, where the sketch  is shown in Fig.~\ref{fig2} (b).
Firstly, we deal with the triplet state, where the spins in Fe$_1$ and Fe$_2$  are parallel to each other ($S_z=1$). 
Assuming $t_{\alpha\sigma/\beta\tau}$ is the hopping integral between the $d_{\alpha}$/$d_{\beta}$ ($\alpha,\beta\in\{xy,x^2-y^2,xz,yz,z^2\}$) Fe$_{1/2}$ and the $p_{\sigma}$/$p_{\tau}$ ($\sigma,\tau\in\{x,y,z\}$) C, the Hamiltonian  for the triplet state  is written by
\begin{equation}\label{append-1}
    H_{\text{T}} =
\begin{bmatrix}
0 & t_{\alpha\sigma} & t_{\beta\tau} & 0 \\
t_{\alpha\sigma} & \Delta_1 & 0 & t_{\beta\tau} \\
t_{\beta\tau} & 0 & \Delta_2 & t_{\alpha\sigma} \\
0 & t_{\beta\tau} & t_{\alpha\sigma} & \Delta_1+\Delta_2 - J_{\sigma\tau}
\end{bmatrix},
\end{equation}
where $\Delta_{1}=U_d+\Delta_{\alpha\sigma}$ and $\Delta_{2}=U_d+\Delta_{\beta\tau}$.  Notably,  $\Delta_{\alpha\sigma}$ ($\Delta_{\beta\tau}$) means the charge‑transfer energy between $d_{\alpha}$  ($d_{\beta}$)  orbital of Fe$_1$ (Fe$_2$) and $p_\sigma$ ($p_\tau$)  orbital of C, where $U_d$ is the  on‑site Coulomb repulsion energy for $d$‑orbitals.
$J_{\sigma\tau}$ is the   intra-atomic Hund exchange integral between two orthogonal occupied different $p_{\sigma}$ and $p_{\tau}$ orbitals of C.  
To realize downfolding, the Hamiltonian matrix can be decomposed by the the following
sub-blocks~\cite{downfolding-1}
\begin{equation}\label{append-2}
\small
\begin{cases}
 H_{PP}=0,   H_{PQ}=H^\dagger_{QP}=\left[t_{\alpha\sigma}, t_{\beta\tau},0  \right]\\
   H_{QQ} =
\begin{bmatrix}
 \Delta_1 & 0 & t_{\beta\tau} \\
  0 & \Delta_2 & t_{\alpha\sigma} \\
 t_{\beta\tau} & t_{\alpha\sigma} & \Delta_1+\Delta_2 - J_{\sigma\tau}
\end{bmatrix}  \\
\end{cases}.
\normalsize
\end{equation}
 $H_{QQ}$ can be regarded as the summation of a diagonal zeroth-order term $H_0$ and  hopping induced perturbative  term V 
\begin{equation}\label{append-4}
\begin{aligned}
H_0=&\text{diag}\left[\Delta_1, \Delta_2,  \Delta_1+\Delta_2 - J_{\sigma\tau}  \right]  \\
=&\begin{bmatrix}
 \Delta_1 & 0 &0 \\
  0 & \Delta_2 & 0 \\
 0 & 0 & \Delta_1+\Delta_2 - J_{\sigma\tau}
\end{bmatrix},
\end{aligned}
\end{equation}
\begin{equation}\label{append-5}
V=\begin{bmatrix}
 0 & 0 & t_{\beta\tau} \\
  0 & 0 & t_{\alpha\sigma} \\
 t_{\beta\tau} & t_{\alpha\sigma} &0
\end{bmatrix}.
\end{equation}
Within the low-energy approximation ($\varepsilon\approx0$), the  zeroth-order propagator is rewritten by
\begin{equation}\label{append-6}
\begin{aligned}
 G_0 &= (\varepsilon I - H_0)^{-1} \approx -H_0^{-1}   
\end{aligned}.
\end{equation}

Expanding the inverse matrix as a Born series, only the even-order terms are non-zero, while the odd-order hopping processes cannot return   to the ground state~\cite{Anderson-1950}.
The second-order term is a single virtual excitation. In such process, one electron  hops from the occupied p orbital of C to the partially-occupied d orbital of Fe and then returns back, which not only leads to lower energy but also preserves  the spin independence. 
More important, the second-order contribution is identical for singlet and triplet configurations,  since it cannot produce any spin splitting or effective exchange coupling.
Furthermore, the  fourth-order term corresponds to two successive virtual excitations.  
In this process,  two electrons  correlates with each other through intermediate states and introduces spin dependence via the exchange energy in the denominator. 
The fourth-order term can  generate the magnetic interaction. 
Therefore, in the expansion, the fourth-order term is the lowest-order process that can generate the superexchange coupling. Neglecting the high order, the effective Hamiltonian  for the triplet  state is expressed by
\begin{equation}\label{append-7}
\begin{aligned}
H_{\text{T}} =& H_{PP} + H_{PQ} \cdot (\varepsilon I - H_{QQ})^{-1} \cdot H_{QP}\\
\approx &  H_{PP} +  H_{PQ} G_0 H_{QP}+
H_{PQ} G_0 V G_0 V G_0 H_{QP}
\end{aligned}.
\end{equation}
Easily, the energy for the triplet state is achieved by
\begin{equation}\label{append-7-sub}
   E_T=-\frac{t_{\alpha\sigma}^2}{\Delta_1} - \frac{t_{\beta\tau}^2}{\Delta_2} - \frac{t_{\alpha\sigma}^2 t_{\beta\tau}^2 (\Delta_1+\Delta_2)^2}{\Delta_1^2 \Delta_2^2 \cdot (\Delta_1+\Delta_2-J_{\sigma\tau})}  .
\end{equation}

As to the  singlet state, the spins of Fe$_1$ and Fe$_2$  have the opposite directions ($S_z=0$). 
The  procedure of the singlet state can be in analogy with that of the triplet state.
Similar to the triplet state, the Hamiltonian is written by
\begin{equation}\label{append-8}
\small
H_{\mathrm{S}}=
\begin{bmatrix}
0 & 0 & t_{\alpha\sigma} & 0 & t_{\beta\tau} & 0 & 0 & 0 \\
0 & 0 & 0 & t_{\alpha\sigma} & 0 & t_{\beta\tau} & 0 & 0 \\
t_{\alpha\sigma} & 0 & \Delta_1 & 0 & 0 & 0 & t_{\beta\tau} & 0 \\
0 & t_{\alpha\sigma} & 0 & \Delta_1 & 0 & 0 & 0 & t_{\beta\tau} \\
t_{\beta\tau} & 0 & 0 & 0 & \Delta_2 & 0 & t_{\alpha\sigma} & 0 \\
0 & t_{\beta\tau} & 0 & 0 & 0 & \Delta_2 & 0 & t_{\alpha\sigma} \\
0 & 0 & t_{\beta\tau} & 0 & t_{\alpha\sigma} & 0 & \Delta_1+\Delta_2 & -J_{\sigma\tau} \\
0 & 0 & 0 & t_{\beta\tau} & 0 & t_{\alpha\sigma} & -J_{\sigma\tau} & \Delta_1+\Delta_2
\end{bmatrix}.
\normalsize
\end{equation}
Now, the downfolding is achieved in two steps, $i.e.$  fold the R-space into the Q-space, then fold the Q-space into the P-space.
The Hamiltonian matrix can be decomposed by the  the following sub-blocks
\begin{equation}\label{append-9}
\small
\begin{cases}
H_{PP}=
\begin{bmatrix}
0&0\\0&0
\end{bmatrix}, \quad H_{QP}=H_{PQ}=\begin{bmatrix}t_{\alpha\sigma} & 0 & t_{\beta\tau} & 0 
\\ 0 & t_{\alpha\sigma} & 0 & t_{\beta\tau}\end{bmatrix}^\dagger,   \\
H_{QQ} =\textbf{diag}\left[\Delta_1,  \Delta_1,\Delta_2, \Delta_2  \right],\quad H_{QR} = H_{RQ}=
\begin{bmatrix}t_{\beta\tau} & 0 & t_{\alpha\sigma} & 0 \\ 
0 & t_{\beta\tau} & 0 & t_{\alpha\sigma}\end{bmatrix}^\dagger,\\
H_{RR} =
\begin{bmatrix}
\Delta_1+\Delta_2 & -J_{\sigma\tau} \\
-J_{\sigma\tau} & \Delta_1+\Delta_2
\end{bmatrix}.
\end{cases}
\normalsize
\end{equation}
Integrating out the R-subspace at energy \(\varepsilon\),  the effective Hamiltonian for the Q-subspace is given by~\cite{downfolding-1}
\begin{equation}\label{append-10}
\widetilde{H}_{QQ} = H_{QQ} + H_{QR}\left(\varepsilon I - H_{RR}\right)^{-1} H_{RQ}.
\end{equation}
Integrating out the Q-subspace,  we can obtain the effective Hamiltonian  in the P-space
\begin{equation}\label{append-11}
\small
\begin{aligned}
H_{\text{eff}}=& H_{PP} + H_{PQ}\left(\varepsilon I - \widetilde{H}_{QQ}\right)^{-1} H_{QP}    \\
=&H_{PP} + H_{PQ}\left[\varepsilon I - H_{QQ} - H_{QR}\left(\varepsilon I - H_{RR}\right)^{-1}H_{RQ}\right]^{-1} H_{QP}
\end{aligned}.
\normalsize
\end{equation}
Considering the low-energy limit ($\varepsilon\approx0$) and $H_{PP}=0$, the  zeroth-order propagator of the singlet state is rewritten by
\begin{equation}\label{append-12}
\small
\begin{aligned}
{G}^R_{0}\approx  - H^{-1}_{RR} , \quad {G}^Q_{0} \approx -{H}^{-1}_{QQ} =    -\left( {H}_{QQ} - H_{QR}  H^{-1}_{RR}H_{RQ}\right)^{-1} 
\end{aligned}.
\normalsize
\end{equation}
Similar to the triplet state,   the Born series for the singlet state should also consider the zeroth, second, and fourth terms.  Thus the effective Hamiltonian is expressed by 
\begin{equation}\label{append-13}
\small
\begin{aligned}
H_{S} \approx &  - H_{PQ}{G}_{0}^Q  H_{QP}
-H_{PQ}\, G_0^{Q}\, H_{QR}\, G_0^{R}\, H_{RQ}\, G_0^{Q}\, H_{QP}\\
=&  \begin{bmatrix}
    \mu &  -\nu\\
    -\nu  &  \mu
\end{bmatrix}
\end{aligned},
\normalsize
\end{equation}
where
\begin{equation}\label{append-14}
    \begin{cases}
 \mu= -\left(\frac{t_{\alpha\sigma}^2}{\Delta_1}+\frac{t_{\beta\tau}^2}{\Delta_2}\right)
-\frac{t_{\alpha\sigma}^2 t_{\beta\tau}^2 (\Delta_1+\Delta_2)^3}
{\left[(\Delta_1+\Delta_2)^2-J_{\sigma\tau}^2\right]\Delta_1^2\Delta_2^2}   \\
\nu= \frac{t_{\alpha\sigma}^2 t_{\beta\tau}^2 (\Delta_1+\Delta_2)^2 J_{\sigma\tau}}
{\left[(\Delta_1+\Delta_2)^2-J_{\sigma\tau}^2\right]\Delta_1^2\Delta_2^2}\\
    \end{cases}.
\end{equation}
After diagonalization of the effective Hamiltonian in Eq.~\eqref{append-13}, the obtained eigenvalues are written by 
\begin{equation}\label{append-15}
\small
    \begin{cases}
E_+=\mu + \nu
=
-\left(\frac{t_{\alpha\sigma}^2}{\Delta_1}+\frac{t_{\beta\tau}^2}{\Delta_2}\right)
-\frac{t_{\alpha\sigma}^2 t_{\beta\tau}^2 (\Delta_1+\Delta_2)^2}
{\Delta_1^2\Delta_2^2 \left(\Delta_1+\Delta_2+J_{\sigma\tau}\right)}
\\
E_-=\mu - \nu =-\frac{t_{\alpha\sigma}^2}{\Delta_1}
-\frac{t_{\beta\tau}^2}{\Delta_2}
-\frac{t_{\alpha\sigma}^2 t_{\beta\tau}^2
(\Delta_1+\Delta_2)^2}
{\Delta_1^2\Delta_2^2
\left(\Delta_1+\Delta_2-J_{\sigma\tau}\right)}
    \end{cases}.
    \normalsize
\end{equation}
Notably, the eigenvalue $E_-$ is exactly the same as the energy for the triplet state in Eq.~\eqref{append-7-sub}. This indicates the spin-rotational invariance of the effective Hamiltonian derived in the $S_z=0$ subspace,   leading to the degeneracy between the 
$S_z=0$ triplet component obtained from the singlet-sector downfolding and 
the $S_z=1$ triplet state. In Eq.~\eqref{append-15}, 
$E_+$ is energy for the singlet state, $i.e.$  $E_S=E_+$. 
Eventually, the superexchange coupling is evaluated in terms of  the energy difference between the triplet ($E_T$) and singlet ($E_+$) states by
\begin{equation}\label{append-18}
J
=
\frac{
2\, t_{\alpha\sigma}^2 t_{\beta\tau}^2
(\Delta_1+\Delta_2)^2 J_{\sigma\tau}
}{
\left[
(\Delta_1+\Delta_2)^2 - J_{\sigma\tau}^2
\right]
\Delta_1^2 \Delta_2^2
}.
\end{equation}

\begin{table}[htbp]
    \centering
    \caption{The amplitude and modulus squared amplitude for the hopping between $p_\sigma$ ($\sigma\in\{ x,y,z\}$) and $d_{\alpha}$ ($\alpha \in\{xy,xz, yz, x^2-y^2,z^2 \}$) orbitals in the units of eV and eV$^2$. }
    \label{tab1}
    \begin{tabular}{cccccccc}
    \hline\hline
     &  & $d_{xy}$ & $d_{x^2-y^2}$ & $d_{xz}$ & $d_{yz}$ & $d_{z^2}$ \\
    \hline
    \multirow{2}{*}{$p_x$} & t&  0.030 & -1.525 & 0.213 & 1.637 & 0.281 \\
                           &t$^2$  & 0.001 & 2.326 & 0.046 & 2.679 & 0.079 \\
    \multirow{2}{*}{$p_y$} & t  & 1.490 & -0.030 & 1.637 & -1.677 & -0.487 \\
                          &t$^2$ & 2.220 & 0.001 & 2.679 & 2.811 & 0.237 \\
    \multirow{2}{*}{$p_z$}&t & 1.824 & 1.053 & 0.495 & -0.857 & 1.214 \\
                          &t$^2$ & 3.327 & 1.109 & 0.245 & 0.735 & 1.473 \\
    \hline
    \hline
    \end{tabular}

\end{table}

It is noticed that Zhang \textit{et al.}~\cite{zhang-2022-gak} have derived the superexchange mechanism on the basis of the $t_{2g}-e_g$ splitting, which is valid for both regular octahedral and tetrahedral crystal field. This mechanism has been successfully applied in quite a few 2D magnetic materials, such as monolayers MoXY (X=S, Se; Y=Br, I)~\cite{lyusai-2023-prb}, heterostructure CrSBr/Bi~\cite{zhou-2205}, CrCoP$_2$S$_6$~\cite{8cnm-qjc8}  and so on.   
However, under the trigonal‑prismatic crystal‑field,   the $d$ orbital of MXene Fe$_2$C  is split into three parts, namely, the non-degenerated irrep $A_1$ $\{d_{z^2}\}$  as well as two  doubly degenerated  $E$ irreps   \{d$_{xy}$, d$_{x^2-y^2}$\} and \{d$_{yz}$, d$_{xz}$\}.
 Regarding this point, we have developed the superexchange mechanism on the basis of such orbital spiting picture as shown in Fig.~\ref{fig2} (b), where the two virtual hopping processes hold different hopping amplitudes.

The exchange coupling parameters can be  analyzed  in a semi-quantitative  way  using Eq.~\eqref{append-18}.   
The Coulomb repulsion interaction ($U_d$ for the d orbital is set as  $U_d=4.0$ eV~\cite{agapov2024mxene,ufe-dft}.  According to the LDOS of the spin down channel (Fig.~\ref{fig2} (a)), the dominant peaks for the occupied orbitals of degenerated $p_x$-$p_y$ and $p_z$ can be roughly regarded as overlapping  near the energy level of about $-$4.3 eV. Likewise, in the spin down channel, the major hole peaks   for   e$_1$ (d$_{xy}$-d$_{x^2-y^2}$),  e$_2$ (d$_{yz}$-d$_{xz}$)   and d$_{z^2}$ nearly coincide at about 0.7 eV. In this regard, the  charge-transfer energy can be treated as a constant of about 5.0 eV. Thus the terms  $\Delta_{1}\approx \Delta_{2}\approx \Delta=U_d+ \Delta_{\alpha\sigma/\beta\tau}=U_d+ \Delta_{pd}  = 9.0$ eV. The Hund exchange integral for p orbitals of C  is assumed to be $J_{\sigma\tau}=$0.6 eV~\cite{ralchenko2005nist}.
Based on Wannier functions, the obtained hopping integrals between  the p orbitals of C and d orbitals of Fe are shown   in Table~\ref{tab1}. 
Owing to  their  near-orthogonality‌,   some orbital pairs are so weak that they can be neglected for the   square term in exchange coupling in Eq.~\eqref{append-18}, namely,  \(d_{xy}\)--\(p_x\) (\(t^2 = 0.000913 \space \) eV$^2$), \(d_{x^2-y^2}\)--\(p_y\) (\(t^2 = 0.000913\) eV$^2$), \(d_{z^2}\)--\(p_x\) (\(t^2 = 0.079132\) eV$^2$), \(d_{xz}\)--\(p_x\) (\(t^2 = 0.045518\) eV$^2$), \(d_{z^2}\)--\(p_y\) (\(t^2 = 0.237397\) eV$^2$), and \(d_{xz}\)--\(p_z\) (\(t^2 = 0.245071\) eV$^2$).   
To understand superexchange, the remaining task is to identify the orbital pairs responsible for intralayer and interlayer exchange coupling, respectively.    Neglecting the hopping with small modulus-squared amplitude ($t^2$),  the orbital pairs of superexchange path  are shown in Table~\ref{tab2}.

\begin{table}
    \caption{The orbital pair for interlayer and intralayer superexchange paths in Fe$_2$C.  $t^2_{\alpha\sigma/\beta\tau}$ is the modulus squared amplitude for the corresponding orbitals in the virtual hopping process in the unit of eV. Parentheses indicate the chemical‑bond types for the respective hopping.
    $J$ denote the exchange coupling contribution in unit of meV.}
    \centering
    \begin{tabular}{cccc}
    \hline\hline
       Orbital pair  & $t_{\alpha\sigma}^2 $ (type) &  $ t_{\beta\tau}^2$ (type)   & $J$    \\
       \hline
       \multicolumn{4}{c}{Intralayer}\\
     \(d_{yz}-p_x/p_y-d_{yz}\)      & 2.679 ($\pi$) & 2.811 ($\sigma$) & 1.378 \\
\(d_{yz}-p_x/p_y-d_{xz}\)      & 2.679 ($\pi$) & 2.679 ($\pi$) & 1.313 \\
\(d_{yz}-p_x/p_y-d_{xy}\)      & 2.679 ($\pi$) & 2.220 ($\pi$) & 1.088  \\
\(d_{x^2-y^2}-p_x/p_y-d_{yz}\) & 2.326 ($\sigma$) & 2.811 ($\sigma$) & 1.197     \\
\(d_{x^2-y^2}-p_x/p_y-d_{xz}\) & 2.326 ($\sigma$) & 2.679 ($\pi$) & 1.140   \\
\(d_{x^2-y^2}-p_x/p_y-d_{xy}\) & 2.326 ($\sigma$) & 2.220 ($\pi$) & 0.945 \\
          \multicolumn{4}{c}{Interlayer}\\
 \(d_{xy}-p_z/p_x-d_{yz}\)        & 3.327 ($\sigma/\pi$) & 2.679 ($\pi$) & 1.631 \\
\(d_{xy}-p_z/p_x-d_{x^2-y^2}\)   & 3.327 ($\sigma/\pi$) & 2.326 ($\sigma$) & 1.416 \\
\(d_{z^2}-p_z/p_x-d_{yz}\)       & 1.473 ($\sigma$) & 2.679 ($\pi$) & 0.722 \\
\(d_{z^2}-p_z/p_x-d_{x^2-y^2}\)  & 1.473 ($\sigma$) & 2.326 ($\sigma$) & 0.627 \\
\(d_{x^2-y^2}-p_z/p_x-d_{yz}\)   & 1.109 ($\sigma/\pi$) & 2.679 ($\pi$) & 0.544 \\
\(d_{x^2-y^2}-p_z/p_x-d_{x^2-y^2}\) & 1.109 ($\sigma/\pi$) & 2.326 ($\sigma$) & 0.472 \\
\(d_{xy}-p_z/p_y-d_{yz}\)      & 3.327 ($\sigma/\pi$) & 2.811 ($\sigma$) & 1.711 \\
\(d_{xy}-p_z/p_y-d_{xz}\)      & 3.327 ($\sigma/\pi$) & 2.679 ($\pi$) & 1.631 \\
\(d_{xy}-p_z/p_y-d_{xy}\)      & 3.327 ($\sigma/\pi$) & 2.220 ($\pi$) & 1.352 \\
\(d_{z^2}-p_z/p_y-d_{yz}\)     & 1.473 ($\sigma$) & 2.811 ($\sigma$) & 0.758 \\
\(d_{z^2}-p_z/p_y-d_{xz}\)     & 1.473 ($\sigma$) & 2.679 ($\pi$) & 0.722 \\
\(d_{z^2}-p_z/p_y-d_{xy}\)     & 1.473 ($\sigma$) & 2.220 ($\pi$) & 0.598 \\
\(d_{x^2-y^2}-p_z/p_y-d_{yz}\) & 1.109 ($\sigma/\pi$) & 2.811 ($\sigma$) & 0.570 \\
\(d_{x^2-y^2}-p_z/p_y-d_{xz}\) & 1.109 ($\sigma/\pi$) & 2.679 ($\pi$) & 0.544 \\
\(d_{x^2-y^2}-p_z/p_y-d_{xy}\) & 1.109 ($\sigma/\pi$) & 2.220 ($\pi$) & 0.451 \\
\hline\hline
    \end{tabular}   
    \label{tab2}
\end{table}

For the intralayer coupling ($J_{in}$),  the two Fe atoms and the bridging C  atoms are approximately in the same $xy$-plane as shown in Fig.~\ref{fig-1} (a) and (b),  leading to a semi-in-plane superexchange. 
Perpendicular to the $xy$ plane, the $p_z$ orbital has a nearly negligible overlapping with the  in-plane $d$ orbitals.    
On the other hand,  according to Slater-Koster scheme~\cite{slater1954simplified} the $p_x$ and $p_y$ orbitals can both form $\sigma$ and $\pi$ types of bonding with the $d$ orbitals of Fe, since both orbitals extend along the $x$ and $y$  directions.  Therefore, the path of the intralayer superexchange   coupling must involve the  orthogonal in-plane  $p_x$ and $p_y$ orbitals,  forming the path  of  $d_\alpha$-$p_x$/$p_y$-$d_\beta$. 
As shown in Table~\ref{tab2}, there are 6 considerable orbital pairs for intralayer superexchange coupling ($J_{in}$), namely \(d_{\alpha}-p_x/p_y-d_{\beta}\) ($\alpha \in \{yz,x^2-y^2\}$, $\beta\in\{ yz,xz,xy\}$).     
As to NN interlayer exchange coupling ($J_1$),  the Fe atoms locate at the  inversion‑symmetric vertically staggered sites  as shown in Fig.~\ref{fig-1} (a) and (b), which are above and below the Carbon layer. 
To mediate superexchange coupling between the bottom and top Fe, the $p_z$ orbital of the bridge C must participate in the superexchange path.   Only involving the  orbital $p_z$, can the hopping of the Fe-$d$ orbital   be propagated out-of-plane via the bridge C.
Since the superexchange path requires two orthogonal $p$ orbitals, the second orbital of C can be either $p_x$ or $p_y$.  Accordingly, there are 15 dominant orbital pairs   contributing to the  NN interlayer exchange coupling ($J_1$) as compiled in Table~\ref{tab2}. The path for the superexchange \(d_{\alpha}-p_z/p_\sigma-d_{\beta}\) ($\alpha \in\{xy, z^2, x^2-y^2 \}$ and $\sigma \in{x,y}$),  can be classified by two groups based on the selection of  $p_\sigma$ orbital. 
For the group of $p_x$ the terminal orbital satisfies $\beta\in\{yz,x^2-y^2 \}$, whereas  for the group of $p_y$,  $\beta$ is restricted to $\{yz,xz,xy \}$.
The NN interlayer exchange coupling ($J_1=5.425$ meV) is stronger than the intralayer exchange coupling ($J_{in}=4.847$ meV),  although the connecting distances of Fe atoms can be comparable (2.594 \AA \space  for the NN interlayer pair and 2.825 \AA for the intralayer pair). 
This can be partially explained by the number  of the available superexchange paths. In the NN interlayer superexchange coupling path, the ligand $p_z$ orbital can pair with either $p_x$ or  $p_y$ orbitals to  form two sets of orthogonal $p$ orbitals, while the orthogonal $p$ orbitals can only be realized  by a pair of both in-plane orbitals $p_x$ and  $p_y$  in the intralayer superexchange coupling path. 
We have not discussed the mechanism for the NNN interlayer exchange coupling (J$_2$)   since it is an unimportant interaction ($J_2=0.766 meV\ll J_1=5.425 meV$).

Finally, We should emphasize that the total superexchange is not the simple algebraic summation of the contributions from different orbital paths, owing to interference through the shared intermediate states and the possible off-diagonal orbital hopping.
Moreover, the direct exchange and superexchange couplings co-exist in the real materials.  
Nevertheless, the semi-quantitative study has identified the dominant superexchange paths, providing a transparent microscopic picture for the exchange couplings in MXene Fe$_2$C. 

\section*{Conclusions}
In conclusion, we have systematically investigated the topology of magnon bands and underlying superexchange mechanism for ferromagnetic MXene Fe$_2$C.
Based on  the exchange coupling parameters from DFT calculations, the spin wave  dispersions are calculated using the linear spin wave method.  In the magnon band structures, a Dirac point at the $K$ point is identified, which originates from the AB-stacking of Fe above and below C layer. Such topological nontrivial phase at the $K$ point is confirmed by the remarkable Berry curvature, valley Chern numbers of $\pm \frac{1}{2}$, and edge states. 
The C$_{3v}$ rotational   symmetry   imposes constraints on the form of the interlayer exchange couplings, 
 Based on the effective Hamiltonian, the topological nontrivial phase is protected by the rotational C$_{3v}$ symmetry  through its restriction on the exchange couplings. 
The microscopic origin of the exchange couplings is demonstrated by   superexchange model based on perturbation theory. 
Under the trigonal-prismatic crystal field,  the d orbitals of Fe  split into three sets, namely the  non-degenerated irrep $A_1$ \{$d_{z^2}$\}  as well as doubly degenerated  $E$ irreps   \{d$_{xy}$, d$_{x^2-y^2}$\} and \{d$_{yz}$, d$_{xz}$\},      while 2p orbitals of C split into  in-plane degenerated E basis $\{ p_x,p_y \}$  and out-of-plane $A_1$ state $p_z$.  
According to GKA rules,  the  ferromagnetic superexchange proceeds via virtual hopping processes through a pair of orthogonal p orbitals at the bridging C site in the spin down channel. 
The intralayer exchange couplings are driven by the in-plane pair  of $p_x$-$p_y$ , whereas the interlayer coupling arises from  the   out-of-plane $p_z$ orbital pairing with the $p_x$ or $p_y$ orbital.
Comparing with the intralayer exchange coupling, the stronger interlayer exchange coupling arises from  more accessible superexchange paths.   
Due to the relatively simple magnetic interactions,   Fe$_2$C   serves as    an ideal platform to explore magnon valleytronics  and superexchange mechanism.

\begin{acknowledgments} 
The authors gratefully acknowledge the computational time on
the Lichtenberg High Performance Supercomputer. This work
was supported by the National Natural Science Foundation of
China (No. 12404045), the Liaoning Provincial Natural Science
Foundation Project of China (No. 2024-MSLH-358),  the Chongqing Natural Science Foundation 
(No. CSTB2025NSCQ-GPX1028), and the
Science and Technology Research Program of Chongqing Municipal Education Commission (No. KJQN-202400553).
\end{acknowledgments}

\bibliography{apssamp}

\end{document}